\documentclass[onecolumn,prb,aps,showpacs]{revtex4}

\usepackage{graphics}% Include figure files

\usepackage{dcolumn}% Align table columns on decimal point

\usepackage{bm}% bold math

\usepackage{epsfig}

\usepackage{pictex}

\begin{document}

\title{Dynamical properties of a strongly correlated model
for quarter-filled layered organic molecular crystals}
\author{Jaime Merino}

\affiliation{Max-Planck-Institut f\"ur Festk\"orperforschung,
D-70506, Stuttgart, Germany}

\author{Andr\'es Greco}

\affiliation{Facultad de Ciencias Exactas Ingenier\'{\i}a y
Agrimensura e Instituto de F\'{\i}sica Rosario (UNR-CONICET),
Rosario, Argentina}

\author{Ross H. McKenzie}

\affiliation{Department of Physics, University of Queensland,
Brisbane 4072, Australia}

\author{Matteo Calandra}

\affiliation{Laboratoire de Min\'eralogie-Cristallographie, case 115, 4
Place Jussieu, 75252, Paris cedex 05, France}

\date{\today}

\begin{abstract}
The dynamical properties of an extended Hubbard model, which is
relevant to quarter-filled layered organic molecular crystals, are
analyzed. We have computed the dynamical charge correlation function, spectral
density, and optical conductivity using Lanczos diagonalization
and large-$N$ techniques.  As the ratio of the nearest-neighbour Coulomb repulsion, $V$,
to the hopping integral, $t$, increases there is
a transition from a metallic phase to a charge ordered phase.
Dynamical properties
close to the ordering transition are found to differ from the ones expected in
a conventional metal. Large-$N$ calculations
display an enhancement of spectral weight at low frequencies
as the system is driven closer to
the charge ordering transition in agreement with Lanczos calculations.
As $V$ is increased the charge correlation
function displays a plasmon-like mode which, for wavevectors
close to $(\pi,\pi)$, increases in amplitude and softens as the charge ordering
transition is approached.  We propose that inelastic X-ray scattering be used to
detect this mode.  Large-$N$ calculations predict superconductivity with
$d_{xy}$ symmetry close to the ordering
transition.  We find that this is consistent with
Lanczos diagonalisation calculations,
on lattices of 20 sites,
which find that the binding energy
of two holes becomes negative close to the charge ordering transition.
\end{abstract}

\pacs{71.27.+a, 71.10.Fd, 74.70.Kn, 71.45.Lr}

\maketitle

\section{Introduction}

The competition between charge ordered, metallic, and
superconducting  phases  is relevant to a broad range of strongly correlated
electron materials.  For example, in the vanadium bronze
$\beta$-Na$_{0.33}$V$_2$O$_5$, superconductivity appears close to
a charge ordered phase under an applied external
pressure\cite{Yamauchi}.  The appearance of a pseudogap
in oxygenated samples of Nd$_{1.85}$
Ce$_{0.15}$ Cu$_{4+y}$  has been suggested to be
due to charge ordering\cite{Onose}.  Quarter-filled layered
organic materials such as the
BEDT-TTF family of organic molecular crystals\cite{Ishiguro}
 with the $\theta$ and $\beta''$ molecular stacking patterns
 also display a subtle
competition of metallic, insulating, charge ordered, and superconducting
phases\cite{McKenzie}.  Superconductivity in organic compounds
is usually found in close proximity to ordered
insulating phases\cite{Ishiguro}.
For example,
$\kappa$-(BEDT-TTF)$_2$Cu[N(CN)$_2$]Cl
 is an antiferromagnetic
Mott insulator which becomes superconducting under pressure\cite{Jerome}.
Superconducting
$\theta$-(ET)$_2$I$_3$ and
$\beta''$-(BEDT-TTF)$_3$Cl$_2$(H$_2$O)$_2$
are close to charge ordered phases\cite{prlmerino}.
Superconductivity occurs in the
quasi-one-dimensional Bechgaard salts,
TMTSF$_2$X, when a spin-density wave
 is suppressed. It is
then important to understand the connection of the superconducting
state to the nearby ordered phases and analyze the effect of
the fluctuations associated with the ordering transition
on the normal metallic phase.

Several anomalous properties have been observed close to the
charge ordering transition in quarter-filled organic conductors:
(i) Suppression of Drude weight and enhancement of optical spectra
at low frequencies at about 500-1000 cm$^{-1}$  in metallic
$\theta$ [\onlinecite{Tajima,Feng}] $\beta''$ [\onlinecite{Dong}]
and $\alpha$-salts [\onlinecite{Dressel}] at low temperatures.
(ii) The temperature dependence of the resistivity may be different from
Fermi liquid behavior, in particular, the resistivity can increase as
the temperature is decreased just before becoming superconducting
(see the Table in Ref. \onlinecite{prlmerino}).
 Previously we have explored, using slave bosons,
the possibility of superconductivity\cite{prlmerino} and the
metal-insulator transition\cite{Calandra} in the quarter-filled
extended Hubbard model. Here, we concentrate on the dynamical
properties in the metallic phase close to the charge ordering
transition. We find that due to the scattering of electrons from
charge fluctuations with $(\pi,\pi)$ wavevector, dynamical and
transport properties display behavior different from that
expected in a typical metal. For
instance, a strong suppression of quasiparticle weight as well as
enhancement of spectral weight at low but finite frequencies takes place as
the charge ordering transition is approached from the metallic
side. Also we examine the possibility of superconductivity
mediated by short range charge fluctuations close to the
transition using both Lanczos diagonalisation and large-$N$
approaches. We find that superconductivity with d$_{xy}$ symmetry
is possible close to the charge ordering transition.  We note that
the present analysis is
similar in spirit to those that aim
to understand the effect of spin fluctuations on the metallic
phase and the possibility of superconductivity mediated by them
\cite{Chubukov} in high-T$_c$ compounds, $\kappa$-(BEDT-TTF)$_2$X
[\onlinecite{Schmalian}], heavy fermions,\cite{mathur}
 and ruthenates\cite{mazin}.

The paper is organized as follows. In Section \ref{sec1}, we
introduce an extended Hubbard model used to describe the
electronic properties of quarter-filled layered molecular
crystals. We also review the path integral formalism written in
terms of Hubbard operators and the large-$N$ expansion introduced
to compute electronic properties of the model. In Section
\ref{sec2}, we show results for the dynamical charge correlation function,
spectral density, and optical conductivity computed with
Lanczos diagonalisation comparing them with
large-$N$ results. In Section IV we discuss
our results contrasting them with available experimental data
on the quarter-filled organics.
 Section V is devoted to the possibility of having
superconductivity in the model.

\section{Dynamical properties in the U-infinite limit:
large-$N$ approach}
\label{sec1}

We consider an extended Hubbard model at one-quarter filling on a
square lattice. This has been argued to be
the simplest model needed to understand the electronic
properties of the layered molecular crystals with
the $\theta$ and $\beta''$ molecular arrangements
within each layer.\cite{McKenzie} The Hamiltonian is
\begin{eqnarray}
H = &-t& \sum_{<ij>,\sigma} (c^\dagger_{i \sigma} c_{j \sigma} +
c^\dagger_{j \sigma} c_{i \sigma}) + U \sum_{i} n_{i\uparrow}
n_{i\downarrow}
\nonumber \\
&+& V \sum_{<ij>} n_i n_j - \mu \sum_{i \sigma} n_{i \sigma}
\label{ham}
\end{eqnarray}
where $U$  and $V$ are the on-site and the nearest-neighbors
Coulomb repulsion, respectively. $c^\dagger_{i \sigma}$ creates an
electron of spin $\sigma$ at site $i$.  In the limit $U \gg V \gg t$
 the ground state is insulating with a checkerboard
charge ordered pattern \cite{McKenzie}.  For $U \rightarrow \infty$ and $V=0$, the
system is expected to be metallic as it is quarter-filled.
Evaluation of the Drude weight by Lanczos techniques
suggests a  metal-insulator transition takes place at a finite
value of $V_c \approx 2.2 t$ for a sufficiently large
value\cite{Calandra} of $U=10t$.

We now introduce the Hubbard operators
\begin{equation}
X_{i}^{0 \sigma} = (1-{c}_{i \bar{\sigma}}^{\dagger}
c_{i \bar{\sigma}}) c_{i \sigma},
 \ \ \ X_{i}^{\sigma 0}= (X_{i}^{0 \sigma})^{\dagger} ,
\ \  \ X_{i}^{\sigma \sigma'}={c_{i\sigma}^{\dagger}}
{c}_{i \sigma'}.
\end{equation}

The five Hubbard ${\hat X_i}$-operators
 $X_i^{\sigma \sigma'}$ and $X_i^{0 0}$ are boson-like
and the four Hubbard ${\hat X}$-operators $X_i^{\sigma 0}$ and
$X_i^{0 \sigma}$ are fermion-like. The names fermion-like and
boson-like come from the fact that Hubbard operators do not verify
the usual fermionic and bosonic commutation
relations\cite{Hubbard}.

In the $U$-infinite limit, the Hamiltonian (\ref{ham})
 ($t-J-V$ model with $J=0$) can be written in terms of Hubbard operators as
\begin{equation}
H(X) = \sum_{<ij>,\sigma}\;t_{ij}\; X_{i}^{\sigma 0} X_{j}^{0
\sigma} + \sum_{<ij>,\sigma} V_{ij} X_{i}^{\sigma \sigma}
X_{j}^{\bar{\sigma} \bar{\sigma}}
-\mu\sum_{i,\sigma}\;X_{i}^{\sigma \sigma}. \label{hamX}
\end{equation}
where $\mu$ is the chemical potential.  The Hubbard operators in
this limit satisfy the completeness condition
\begin{equation}
X_{i}^{0 0} + \sum_{\sigma} X_{i}^{\sigma \sigma} = 1 ,
\label{constr}
\end{equation}
which is equivalent to imposing
that "double occupancy" at each site is forbidden.

There are two main difficulties in the calculation of physical
quantities using Hamiltonian (\ref{ham}): the complicated
commutation rules of the Hubbard operators \cite{Hubbard} and that
there is no small parameter in the model. A popular method for
handling the former difficulty is to use slave particles. For
instance, within the slave boson method \cite{Kotliar}, the
original fermionic $X^{0 \sigma}$ operator is decoupled as $X^{0
\sigma} = b^{\dagger} f_{\sigma}$, where $b$ and $f$ are usual
boson and fermion operators, respectively. The second difficulty
can be dealt with by using a non-perturbative technique (which we
will use in the present paper) based on a large-$N$ expansion,
where $N$ is the number of electronic degrees of freedom per site
and 1/$N$ is assumed to be a small parameter. At one-quarter
filling (which is the main interest in this paper), we expect the
large-$N$ approach to be a good approximation. This has been shown
in the overdoped regime of high-$T_c$ cuprates\cite{Zeyher}.

Hamiltonian (\ref{ham}) has been treated via large-$N$ in a slave
boson representation in Ref. {\onlinecite{Kotliar} for $V=0$, and
in the context of quarter-filled layered organic superconductors
($V \neq 0$) in Ref. \onlinecite{McKenzie}. Here, we concentrate
on the dynamical properties of Hamiltonian (\ref{ham}), using the
recently developed large-$N$ expansion \cite{Foussats}. This
method is based on a path integral representation of the Hubbard
$X$-operators which is written in terms of Grassmann and usual
bosonic variables associated with fermi-like and boson-like
operators, respectively.  In doing this, additional constraints
are needed to make these field variables behave as Hubbard
operators (satisfying their associated algebra), as they should.
Although this may seem a great complication in the theory, in fact
it avoids introducing any decoupling scheme of the original
Hubbard operators, as in slave boson representations. For
completeness we will summarize the framework used in the
diagrammatic expansion developed in Ref. [\onlinecite{Foussats}].

Our starting point is the partition function $Z$ written in the
Euclidean form
\begin{eqnarray}
Z&=&\int {\cal D}X_{i}^{\alpha \beta}\;
\delta[X_{i}^{0 0} + \sum_{\sigma} X_{i}^{\sigma \sigma}-1]\;
\delta[X_{i}^{\sigma \sigma'} - \frac{X_{i}^{\sigma 0}
X_{i}^{0 \sigma'}}{X_{i}^{0 0}}] \nonumber \\
&\times&({\rm sdet} M_{AB})_{i}^{\frac{1}{2}} \exp\;(- \int d\tau\;L_E(X,
\dot{X}))\;. \label{part}
\end{eqnarray}

The Euclidean Lagrangian $L_E(X,\dot{X})$ in (\ref{part}) is
\begin{eqnarray}
L_E(X, \dot{X}) =  \frac{1}{2}
\sum_{i, \sigma}\frac{({\dot{X_{i}}}^{0 \sigma}\;X_{i}^{\sigma 0}
+ {\dot{X_{i}}}^{\sigma 0}\;
X_{i}^{0 \sigma})}
 {X_{i}^{0 0}}
+ H(X)\;. \label{lagr}
\end{eqnarray}

It is worth noting at this point that the path integral
representation of the partition function \ref{part}, looks
different to that usually found in other solid state problems. The
measure of the integral contains additional constraints  as well
as a superdeterminant, $( {\rm sdet} M_{AB})_{i}^{\frac{1}{2}}$.  Also the
kinetic term of the Lagrangian (\ref{lagr}) is non-polynomial. The
determinant reads
\begin{equation}
({\rm sdet} M_{AB})_{i}^{\frac{1}{2}}=1/{1 \over (-X^{00})^2},
\end{equation}
and is formed by all the constraints of the theory. Note that
$({\rm sdet} M_{AB})_{i}^{\frac{1}{2}}$ is not proportional to
$(-X^{00})^2$, because the theory is constrained in a
supersymmetric sense where boson and fermion determinants must be
treated in different ways (see Ref[\onlinecite{Foussats}] for more
details about the path integral formalism for Hubbard operators).
The constraints appearing in the theory are necessary in order to
recover the correct algebra of the original Hubbard operators. In
Eq. (\ref{lagreff}) we show how to treat this determinant through
the use of a large-$N$ expansion.

We now discuss the main steps needed to introduce a large-$N$
expansion of the partition function (\ref{part}). First, we
integrate over the boson variables $X^{\sigma \sigma'}$ using the
second $\delta$-function in (\ref{part}).  We extend the spin
index $\sigma=\pm$, to a new index $p$ running from $1$ to $N$. In
order to get a finite theory in the $N\rightarrow \infty$ limit,
we re-scale the hopping $t_{ij}$ to $t_{ij}/N$ and $V_{i j}$ to
$V_{i j}/N$.  In doing so, note that $t_{ij}/N$ (rather than
$t_{ij}$), should be fitted to band structure calculations. The
completeness condition is enforced by exponentiating: $X_{i}^{0 0}
+ \sum_{p} X_{i}^{p p} = N/2 $, with the help of Lagrangian
multipliers $\lambda_i$. We write the boson fields in terms of
static mean-field values, $(r_0, \lambda_0)$ and dynamic
fluctuations
\begin{eqnarray}
X_{i}^{0 0} &=& N r_{0}(1 + \delta R_{i})
\nonumber \\
\lambda_{i} &=&\lambda_{0}+ \delta{\lambda_{i}},
\label{hubbop}
\end{eqnarray}
and, finally, we make the following change of variables
\begin{eqnarray}
f^{+}_{i p} &=& \frac{1}{\sqrt{N r_{o}}}X_{i}^{p 0}
\nonumber \\
f_{i p} &=& \frac{1}{\sqrt{N r_{o}}}\;X_{i}^{0 p},
\label{fermoper}
\end{eqnarray}
where $f^{+}_{i p}$ and $f_{i p}$ are Grassmann variables.

Introducing the above change of variables (Eq. (\ref{hubbop}) and
Eq. (\ref{fermoper})) into Eq. (\ref{lagr}) and, after expanding the
denominator appearing in (\ref{lagr}), we arrive at the following
effective Lagrangian:
\begin{eqnarray}
&L_{eff}&=-\frac{1}{2}\sum_{i,p}^{N}\left(\dot{f_{i p}}f^{+}_{i p}
+ \dot{f^{+}_{i p}}f_{i p}\right) (1 - \delta R_{i} + \delta
R_{i}^{2}) \nonumber \\
&+&\sum_{i,j,p}^{N}\;t_{ij} r_{o} f^{+}_{i p}f_{j p}+
\sum_{i,j}^{N}\;V_{ij} r_{o}^2 \delta R_{i} \delta R_{j} -
\mu\;\sum_{i,p}\;f^{+}_{i p}f_{i p}
(1 - \delta R_{i} + \delta R_{i}^{2}) \nonumber \\
&+&N\;r_{0}\;\sum_{i}\delta{\lambda_{i}}\;\delta R_{i}
+\sum_{i,p}
f^{+}_{i p}f_{i p}(1 - \delta R_{i} + \delta R_{i}^{2})\;
\delta{\lambda_{i}} \nonumber \\
&-&\sum_{i p}\;
{\bf {\cal Z}}_{i p}^{\dag}\left(1- \delta R_i+ \delta R^2_i\right)
{\bf {\cal Z}}_{i p},
\label{lagreff}
\end{eqnarray}
where $\lambda_0$ has been absorbed in the chemical potential $\mu
\rightarrow \mu-\lambda_0$ and all constant and linear terms in
the fields have been dropped.  The path integral representation of
$({\rm sdet} M_{AB})^{\frac{1}{2}}$, written in terms of the
$N$-component boson ghost fields, \cite{van} ${\bf{\cal Z}}_{p}$,
leads to the last term of Lagrangian (\ref{lagreff}). Note that
all the complications arising from the Hubbard algebra have been
translated to an effective theory of fermions interacting with
bosons. Indeed, the interaction terms appearing in the effective
lagrangian (\ref{lagreff}) are generated solely by the Hubbard algebra (apart from the no
double occupancy constraint) and are not present in the original
hamiltonian (\ref{hamX}), which is quadratic in the Hubbard
operators.

In the above expansion we have only retained the first non-trivial
terms that couple the fermionic and bosonic modes.  In order to
have a systematic scheme to classify and deal with these
interaction terms we introduce a set of Feynman rules in powers of
$1/N$ \cite{Foussats}.  These will help us to determine, for
instance, that the terms retained in the effective Lagrangian (\ref{lagreff})
correspond to expanding through O($1/N$) in the large-$N$ expansion.
The Feynman rules needed to carry out this project can be summarized as follows:

{\it (i) Propagators}: We associate with the two component $\delta
X^{a} = (\delta R\;,\;\delta{\lambda})$ boson field, the bare
propagator $D_0$
\begin{eqnarray}
D^{-1}_{(0) ab}({\bf q},\nu_n)=N
\left(
\begin{array}{cc}
4 Vr^2_0(\cos(q_x)+\cos(q_y))    & r_{0}  \\
r_{0}      & 0
\end{array}
\right)
\end{eqnarray}
\noindent which is represented by a dashed line in Fig. \ref{figvertex}
connecting two generic components $a$ and $b$. $q$ and
$\nu_{n}$ are the momentum and the Matsubara frequency of the
boson fields, respectively.

The bare propagator of the $N$-component fermion field $f_{p}$ reads
\begin{eqnarray}
G_{(0)pp'}({\bf k}, \omega_n) = - \frac{\delta_{pp'}}{i\omega_n -
(\varepsilon_{{\bf k}} - \mu )}
\label{fermprop}
\end{eqnarray}
\noindent which is represented by a solid line in
Fig.\ref{figvertex} connecting two generic components $p$ and
$p'$. The electron dispersion relation appearing in Eq.
(\ref{fermprop}) is the one associated with the original fermions
renormalized by the interaction: $\varepsilon_{{\bf k}} = -2tr_{o}
(\cos(k_x)+\cos(k_y))$, with $t$ the hopping between nearest
neighbors sites on the square lattice. The quantities $k$ and
$\omega_{n}$ are the momentum and the fermionic Matsubara
frequencies of the fermion field, respectively.

We associate with the $N$-component bare ghost field ${\bf {\cal
Z}}_{p}$ the propagator
\begin{equation}
{\cal D}_{pp'} = - \delta_{pp'}
\end{equation}
\noindent which is represented by a dotted line in Fig.
\ref{figvertex} connecting two generic components
p and p'.\\

{\it (ii) Vertices}: The expressions of the different three-leg
and four-leg vertices are
\begin{eqnarray}
\Lambda^{pp'}_{a} = -\; \left(\frac{i}{2}(\omega + \omega') +
\mu\;;\;1\right)\;\delta^{pp'}\, \label{vertexa}
\end{eqnarray}
\noindent representing the interaction between two fermions and
one boson (see Fig.\ref{figvertex}(a));
\begin{eqnarray}
\Lambda^{pp'}_{ab} = -
\left(
\begin{array}{cc}
- \frac{i}{2} (\omega + \omega') - \mu   &- \frac{1}{2}      \\
- \frac{1}{2}                                          &0
\end{array}
\right)\delta^{pp'}, \label{vertexb}
\end{eqnarray}
\noindent representing the interaction between two fermions and
two bosons (see Fig. \ref{figvertex}(b));
\begin{eqnarray}
\Gamma^{a}_{pp'} = (-1) (\delta_{pp'}\;,\;0),
\end{eqnarray}
\noindent representing the interaction between two ghosts and one
boson (Fig. \ref{figvertex}(c)); and
\begin{eqnarray}
{\Gamma}^{ab}_{pp'} = (-1)
\left(
\begin{array}{cc}
-1  &0  \\
0  &0
\end{array}
\right) \delta_{pp'}\, \nonumber\\
\end{eqnarray}
\noindent representing the interaction between two bosons and two
ghosts (Fig. \ref{figvertex}(d)).  Each vertex conserves momentum
and energy, as it should.
\begin{figure}
\begin{center}
\epsfig{file=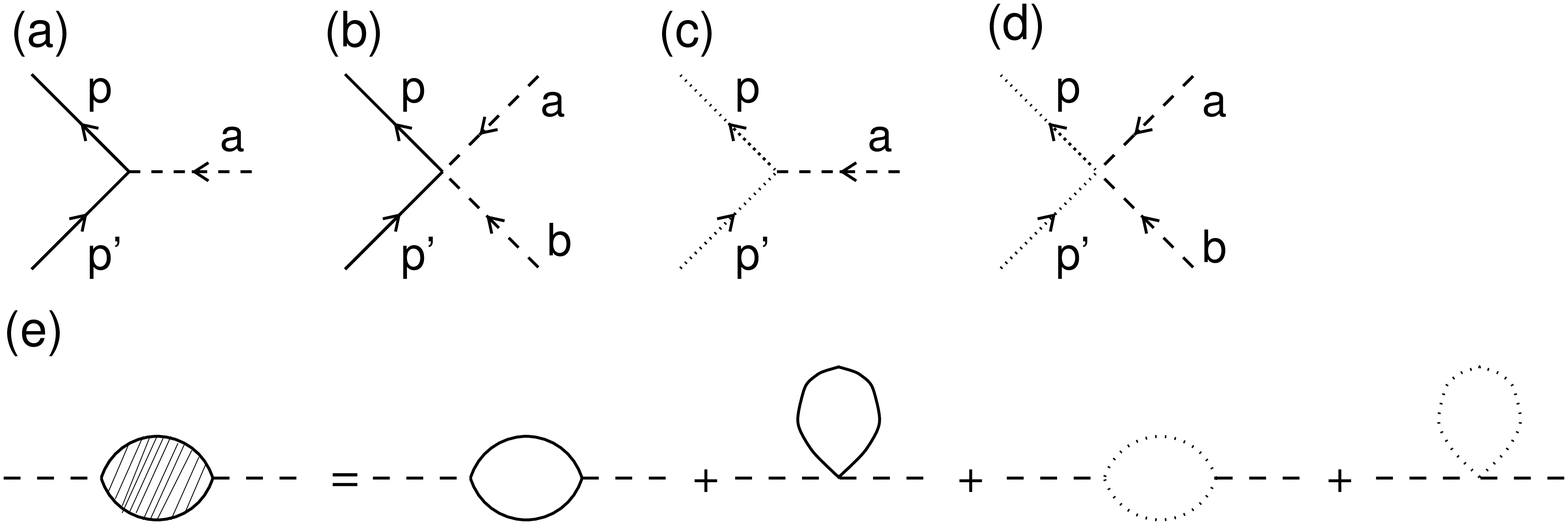,width=8.0cm,angle=0}
\end{center}
\caption{Feynman diagrams in the large-$N$ expansion of the
Hubbard operator theory. Solid lines represent fermions which are
related to the electrons. Dashed lines represent bosons which are
related to charge fluctuations. Dotted lines represent ghosts
which are not physical but related to the constraints appearing in
the theory which enforce that fermions satisfy the Hubbard
operator algebra. (a) to (d) show the types of
vertex which occur up to order $O(1/N)$. (a) is the vertex between
two fermions and one boson. (b) is the vertex between
 two fermions and two bosons. (c) is the vertex between two
ghosts and one boson. (d) shows the vertex between
 two bosons and two ghosts. (e)
represents the sum of all one-loop diagrams contributing to the
irreducible boson self energy which is of O($1/N$). } \label{figvertex}
\end{figure}

In the lowest order of the expansion $N = \infty$, we have the
original fermions renormalized by the interaction,
$\varepsilon_{{\bf k}} = -2tr_{o} (\cos(k_x)+\cos(k_y))$. For a given
value of $\mu$, $r_{0}$ must be determined self-consistently.
For instance, $r_{0}$ is equal to $\delta/2$
(where $\delta$ is the hole doping away from half-filling)
from (\ref{hubbop}) and the completeness condition (\ref{constr}).

The path integral (\ref{part}) is written in terms of the original
$X$-operators without having to introduce slave particles.
Eq.(\ref{part}) is analogue to the path integral used for the
Heisenberg model where, using $SU(2)$ coherent states, the measure
can be written\cite{Fradkin} in terms of the spin $\vec{S}$. There
is, however, an extra price we have to pay if we work with the
original Hubbard operators. For instance, we need to introduce a
new constraint (the second delta function in Eq. (\ref{part})) and
the determinant ($ {\rm sdet} M_{AB}$) of the matrix formed by the
constraints appearing in the theory \cite{Foussats}. In spite of
these "apparent" complications our formulation is very flexible in
calculating the physical quantities of interest, as it will be
shown below.

In summary, we have developed a diagrammatic technique appropriate
for a large-$N$ expansion along the lines of the large-$N$
expansion developed in quantum field theory. Hence, from the order
of the propagators and vertices, we can determine the order of the
diagram contribution.

To conclude this section we make contact with closely related
approaches such as slave boson formulations.  In contrast to slave
boson theories: (a) Greens functions are calculated in terms of
the original Hubbard operators, (b) fermions, $f_{ip}$, appearing
in the theory are proportional to the Fermi-like $X$-operator
$X^{op}$ (see (\ref{hubbop})) to all orders in the $1/N$
expansion; not only to leading order\cite{Wang}, (c) as our path
integral is written in terms of $X$-operators we do not need to
introduce {\it a priori} any decoupling scheme, and (d) $r_0$ is
the mean value of $X^{00}$ which is a real field associated with
the number of holes (see Eq. (\ref{hubbop})) and not with the
number of holons. At leading order ($N \rightarrow \infty$ or
$O(1)$) and $V=0$, our formalism is equivalent to slave boson
approaches. However, at the next to leading order ($O(1/N)$),
(which is necessary to calculate one-electron properties such as
the electron self energy $\Sigma({\bf k},\omega)$ and the electron
spectral function $A({\bf k},\omega)$), the two formulations do
not coincide.  The differences between the two formulations are
not yet completely established.  Our theory has the {\it
significant advantage} that it does not require the introduction
of gauge fields like in slave boson approaches. Hence, through
order $O(1/N)$ we do not need to take care of gauge fluctuations
nor Bose condensation (note that Eq. \ref{hubbop} does not mean
Bose condensation).  This is important because for the doped
Hubbard model the gauge fluctuations are known to significantly
change the physics\cite{ioffe}.  Careful numerical work will
determine the improvements of the present approach with respect to
slave boson formulations.

\section{Dynamical properties of the metallic phase close
to the charge ordering transition}
\label{sec2}

In this section, we analyze using large-$N$ and Lanczos techniques
the influence of the charge ordering transition on the dynamical
properties of the normal metallic phase.

\subsection{Charge response}

The dynamical electronic density-density response function
can be written in terms of Hubbard operators.  We define the
retarded density-density, Green's function as
\begin{equation}
{\tilde{D}}_{ij}(\tau) =\frac{1}{N} \sum_{pq} <T_{\tau} X^{pp}_i(\tau)
X^{qq}_j(0)>.
\end{equation}

From $\sum_q X^{qq}_i=N/2- X^{00}_i$ and (\ref{hubbop}) we find,
after Fourier transforming,
\begin{equation}
\tilde{D}({\bf q},\nu_n)=-N {(\frac{\delta}{2})}^2 D_{RR}({\bf
q},\nu_n). \label{bosonp}
\end{equation}

Here $D_{RR}({\bf q},\nu_n)$ is the $(R,R)$ component of the boson
propagator. This is the only physical component of the boson
propagator and encodes the charge fluctuations occurring in the
system. Other components of the boson propagator such as the $(\lambda, R)$ or
$(\lambda, \lambda)$ contain the nonphysical field $\lambda$ which
are introduced to enforce the no double-occupancy constraint. Unlike in slave boson
theories, the $(R,R)$ component used here is associated directly
with the charge and not with a fictitious bosonic field (holon).

Through $O(1/N)$ the boson propagator consists of the bare boson
propagator $D_{(0)}$ (which is of order $O(1/N)$) renormalized by
a RPA-type series of electronic bubbles.  The irreducible boson
self-energy components, $\Pi_{ab}$, are obtained (through order
$1/N$) from the summation of all the contributions corresponding
to the one-loop diagrams shown in Fig. \ref{figvertex}(e).

The last two diagrams appearing in Fig. \ref{figvertex}(e)
involving ghost fields are very important. It is possible to show
that these two diagrams exactly cancel the infinities, due to the
frequency dependence of our vertices, of the two first diagrams
appearing in Fig. \ref{figvertex}(e). Ghost fields interact only
with the boson fields as can be seen from Fig. \ref{figvertex}(c)
and Fig. \ref{figvertex}(d). Summarizing, the only role of ghost
fields, through order $1/N$, is to cancel infinities in the boson
self-energy $\Pi_{ab}$ arising from the frequency dependence of
our vertices (\ref{vertexa}) and (\ref{vertexb}).

Using our Feynman rules, we can now write out explicitly each of
the components of the boson self-energy $\Pi_{ab}$
\begin{eqnarray}
\Pi_{RR}({\bf q}, \nu_{n}) &=& -\frac{N}{N_{s}}\;\frac{1}{4}
\sum_{{\bf k}}\;\left[ 2\;n_{F}(\varepsilon_{{\bf k}} - \mu)
(\varepsilon_{ \bf{k +
q}} - \varepsilon_{{\bf k}})
\right.\nonumber\\
& + &\left.(\varepsilon_{{\bf k + q}} + \varepsilon_{{\bf k}})^{2}\;
\frac{[n_{F}(\varepsilon_{{\bf k + q}} - \mu) - n_{F}(\varepsilon_{{\bf k}} -
\mu)]} {- i \nu_{n} + \varepsilon_{{\bf k + q}} -
\varepsilon_{{\bf k}}}\right]\;,
\end{eqnarray}

\begin{eqnarray}
\Pi_{\lambda R}({\bf q}, \nu_{n}) &=& - \frac{N}{N_{s} }\;\frac{1}{2}
\sum_{{\bf k}}\;(\varepsilon_{{\bf k + q}} + \varepsilon_{{\bf k}}) \nonumber \\
&\times & \frac{[n_{F}(\varepsilon_{ {\bf k + q}} - \mu) -
n_{F}(\varepsilon_{{\bf k}} - \mu)]} {- i \nu_{n} + \varepsilon_{ {\bf k +
q} } - \varepsilon_{{\bf k}}}
\end{eqnarray}
\noindent and,

\begin{eqnarray}
\Pi_{\lambda\lambda}({\bf q}, \nu_{n}) = - \frac{N}{N_{s}}\;
\sum_{{\bf k}}\;\frac{[n_{F}(\varepsilon_{{\bf k + q}} - \mu) -
n_{F}(\varepsilon_{{\bf k}} - \mu)]} {- i \nu_{n} + \varepsilon_{{\bf k +
q}} - \varepsilon_{{\bf k} }}\;.
\end{eqnarray}
where $N_s$ is the number of sites of the system.

From Dyson's equation and $\Pi_{ab}$ the dressed components of the
boson propagator, $D_{ab}$, can be found:
\begin{equation}
(D_{ab})^{-1} = (D_{(0) ab})^{-1} - \Pi_{ab}
\label{Dab}
\end{equation}
$D_{ab}$ may contain collective excitations such as zero sound
\cite{Wang}.

In order to look at charge ordering instabilities induced by the
intersite Coulomb interaction, $V$, we have calculated the static
charge susceptibility $\tilde{D}({\bf q},\nu_n=0)$ for different
${\bf q}$ vectors on the Brillouin zone (BZ). At one-quarter
filling ($\delta=0.5$) the corresponding chemical potential is
$\mu=-0.360t$ in the limit $N \rightarrow \infty $. We find that
the static susceptibility diverges at the wavevector ${\bf
q_c}=(\pi,\pi)$ for $V=V_c \approx 0.65t$ signalling the
instability to a checkerboard charge density wave (CDW). The value
of $V_c$ is slightly smaller than the one previously found using
slave bosons\cite{McKenzie}, $V_c \approx 0.69t$. This is because
of the decoupling of the electron operators introduced within
slave bosons to treat the intersite interaction term, $V n_i n_j$
which is not needed (due to the use of Hubbard operators)
in the present large-$N$ approach. For comparison,
recent exact diagonalization calculations\cite{Calandra} give a
critical value for the metal-insulator transition driven by $V$ at
about $V_c \approx 2 t$ for $U = 20t$. The large difference in
$V_c$ between large-$N$ and Lanczos diagonalization calculations
can be attributed to the strong renormalization of the bare band
(given by $r_0=\delta/2$) which appears in large-$N$ approaches at
$O(1)$. Introducing higher order terms in the $1/N$-expansion may
give larger values of $V_c$, in closer agreement with Lanczos
calculations.

In Fig. \ref{figcharge} we show the evolution of $-{\rm
Im}\tilde{D}({\bf q_c},\nu)$ as the system is driven close to the
charge ordering instability, $V< V_c$, for the wavevector: ${\bf
q_c}=(\pi,\pi)$. The intersite Coulomb repulsion softens the
plasmon mode at ${\bf q_c}$ which appears for $U \rightarrow
\infty $ and $V=0$ and, at the same time, increases its weight.
At wavevectors far from ${\bf q_c}$ the plasmon mode shows up as a
peak located at frequencies of about $t$ which carries small
weight and is barely influenced by $V$. Because the mode at
$(\pi,\pi)$ is plasmon-like it can be detected, in principle, with
electron energy loss scattering (EELS)\cite{Fink} or inelastic
X-ray scattering \cite{Hasan}. With EELS one is able to obtain
information on the electronic properties of the system at a given
energy and wavevector, so that, for instance, the dispersion
relation of the mode can be mapped out. A more appropriate way of
detecting the plasmon-like mode is by using inelastic X-ray
scattering, which provides a direct probe of the dynamical charge
correlation function and has been succesfully applied to one- and
two-dimensional Mott-Hubbard systems\cite{Hasan}.

In order to compare with large-$N$ we compute, with Lanczos
diagonalization, the spectral decomposition of the charge
correlation function
\begin{equation}
C({\bf q}, \nu)=\sum_{m} |\langle m | N_{\bf
q} | 0 \rangle|^2 \delta (\nu-(E_m-E_0))
\end{equation}
where $N_{\bf q}= 1/\sqrt{L}\sum_i e^{i {\bf q} {\bf R_i}} (c^+_i
c_i- \langle c^+_i c_i \rangle ) $. $E_m$ and $E_0$ denote the
excited and ground state energies of the system, respectively. $L$
is the number of sites in the cluster.  Note that $C({\bf q},
\nu)$ can be compared to $-{\rm Im}\tilde{D}({\bf q_c},\nu)$ as
they have equivalent definitions. Of course, attention must be
paid to the fact that we are comparing calculations of the charge
susceptibility on an infinite system with calculations on a $L=16$
cluster.  Indeed, we find that $C({\bf q_c}, \nu)$ is in rather
good agreement with $-{\rm Im}\tilde{D}({\bf q_c},\nu)$ (see Fig.
\ref{figcharge}), both displaying similar softening and increase
in amplitude of the plasmon mode at $(\pi,\pi)$ close to the
charge ordering transition.

The imaginary part of the charge correlation close to the charge
ordering wavevector: ${\bf q} \rightarrow {\bf q_c}$
can be fitted to the
following RPA form \cite{Castellani}
\begin{equation}
-{\rm Im} \tilde{D}({\bf q}, \nu)= A {\nu \over \nu^2+ \omega_{\bf q}^2}
\label{sing}
\end{equation}
where $\omega_{\bf q}=\omega_0+ C({\bf q}- {\bf q_c})^2$, where $A$ and $C$
are constants.  $\omega_0$ gives the position of the peak appearing in the
charge correlation function at $(\pi,\pi)$ for
different $V$'s and goes to zero as $V \rightarrow V_c$, measuring
the proximity of the system to the charge ordering transition (see inset
of Fig. \ref{figcharge}).  We
note that the overall behavior of the charge susceptibility
is analogous to the one
of the spin susceptibility in nearly
antiferromagnetic metals\cite{Hlubina,Yanase}.
\begin{figure}
\resizebox{!}{10cm}{\includegraphics{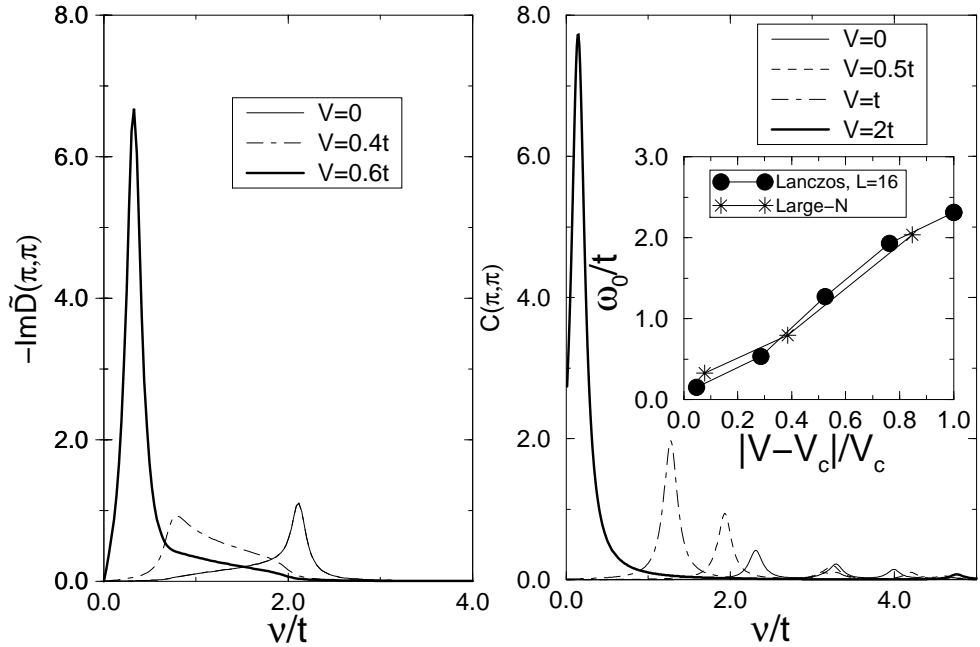}} \caption{ The
softening of the plasmon-like mode at the wavevector
 ${\bf q_c }=(\pi,\pi)$ as the system is driven closer to the checkerboard charge
ordering transition. The frequency dependence of the charge
correlation function is shown for several different values of
$V/t$. The right and left panel show results obtained using
Lanczos diagonalization on $L=16$ site clusters ($U=20t$) and
large-$N$ theory, respectively. A Lorentzian broadening of
$\eta=0.1t$ is used in the calculations. Only for wavevectors
close to or at $(\pi,\pi)$, the softening of the plasmon mode is
observed as a consequence of the proximity of the system to a
checkerboard charge ordering transition. Calculations of dynamical
properties using large-$N$ theory at O($1/N$), which couples the
electrons to the short range charge fluctuations associated with
this transition, and Lanczos diagonalization, suggest that this
plasmon mode is responsible for the 'unconventional' behavior of
dynamical properties. The inset compares the position of the
plasmon peak at $(\pi,\pi)$ computed from Lanczos and large-$N$
approaches. } \label{figcharge}
\end{figure}

\subsection{Spectral densities}

In order to calculate spectral densities, we first discuss the
evaluation of the self-energy corrections to the bare fermion
propagator (\ref{fermprop}), which occur at $O(1/N)$. Using our
Feynman rules there are two diagrams, shown in Fig.
\ref{figfeynman}(a), contributing to the self energy to $O(1/N)$.
The analytical expression for these two diagrams reads:
\begin{equation}
\Sigma_{pp}=\sum_{p',p'',a,b} \Lambda^{pp'}_{a} D_{ab} G_{p'p"}
\Lambda^{p"p}_{b} + \sum_{a,b}\Lambda^{pp}_{ab} D_{ab}
\end{equation}
where integration over internal momenta and sum over Matsubara
frequencies is assumed.

Using the spectral representation for the boson fields, $D_{ab}$,
the imaginary part of the self-energy $Im\Sigma$ can be obtained

\begin{figure}
\begin{center}
\epsfig{file=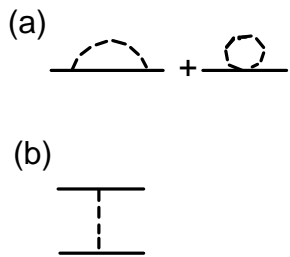,width=8.0cm,angle=0}
\end{center}
\caption{(a) Contributions to the electron self-energy,
$\Sigma({\bf k}, \omega)$, through $O(1/N)$, in the Hubbard
operator theory. The first diagram contains two three-leg vertices
as the ones shown in Fig. \ref{figvertex}(a) and the second one is
formed with one four-leg vertex as shown in Fig.
\ref{figvertex}(b). (b) Contribution to the effective interaction
between quasiparticles, $V_{eff}$, through $O(1/N)$. This
interaction is used in the present work to analyze superconducting
instabilities of the Fermi liquid induced by the charge
fluctuations appearing close to a checkerboard charge ordering
transition induce by $V$.} \label{figfeynman}
\end{figure}
\begin{eqnarray}
{\rm Im}\Sigma({\bf k},\omega)&=&-\frac{1}{N_s} \sum_{\bf q} \{
\frac{1}{4} {\rm Im}(D_{RR}({\bf q}, \omega- \varepsilon_{{\bf
k-q}}))
(\varepsilon_{k-q} +2 \mu
+\omega)^2 \nonumber \\
&+& {\rm Im}(D_{R\lambda}( {\bf q},\omega- \varepsilon_{{\bf
k-q}})) (\varepsilon_{ {\bf k-q}} +2 \mu +\omega)
\nonumber  \\
&+& {\rm Im}(D_{\lambda \lambda}({\bf q},\omega-
\varepsilon_{{\bf k-q}}))\} (n_B (\omega-\varepsilon_{{\bf k-q}})
+n_F(-\varepsilon_{{\bf k-q}}))
\label{iself}
\end{eqnarray}

Note that this self-energy is the one associated with the
propagator $G({\bf k},\omega)$ of the Hubbard $X$-operators. In
contrast, in slave-boson approaches a convolution of the fermion
and boson operators is needed in order to recover the actual
electronic self-energy\cite{Wang}.

\begin{figure}
\resizebox{!}{10cm}{\includegraphics{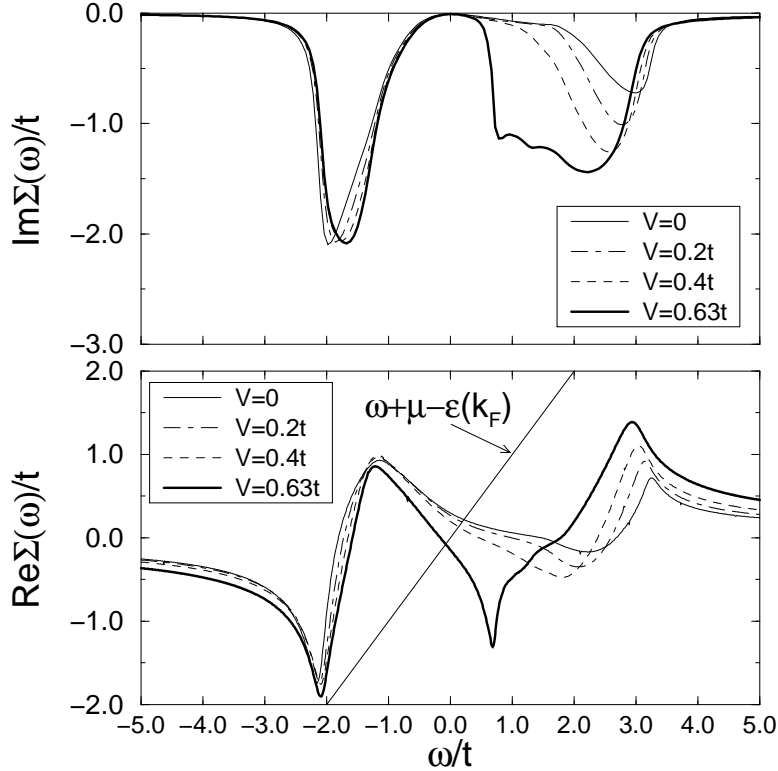}}
\caption{Evolution of the real and imaginary parts of the
self-energy of an electron at the Fermi surface as the system is
driven close to the checkerboard charge ordering transition from
large-$N$ theory. The amplitude of the self-energy is enhanced at
frequencies between $t$ and $3t$ due to the enhancement of
fluctuations associated with $(\pi,\pi)$ short range charge ordering.
The behavior of the self-energy leads to an enhancement of spectral
weight in the spectral density
(see Figs. \ref{figspect}) and an incoherent band  in the
DOS (see Fig. \ref{figdos}) between $\omega=t$ and $3t$ as
we approach the charge ordering transition.
The intersection of the curve of ${\rm Re} \Sigma(\omega)$ versus
$\omega$ with $\omega + \mu - \epsilon({\bf k })$ determines
the quasi-particle peaks in the electronic spectral function.
} \label{figself}
\end{figure}

Fig. \ref{figself} shows the behavior of Im$\Sigma({\bf
k},\omega)$ with increasing $V$ for a wavevector on the Fermi
surface: ${\bf k}=(1.204,1.204)$ (we have used $\eta=0.1t$ in the
analytical continuation).  As we approach $V=V_c$, both the
imaginary and real parts of the self-energy (which from Eq.
(\ref{iself} involves a sum over the full BZ) are enhanced in the
positive range of frequencies $t-3t$ due to the scattering of the
electrons off the checkerboard charge fluctuations. Performing a
Kramers-Kronig transformation on ${\rm Im}\Sigma$, we can obtain
the real part of the self-energy, Re$\Sigma$, which is also
plotted in Fig. \ref{figself}.

From ${\rm Im}\Sigma$ and Re$\Sigma$, we can compute the electron
spectral function $A({\bf k},\omega)=-\frac{1}{\pi}
 {\rm Im}G({\bf k},\omega)$ as
\begin{eqnarray}
A({\bf k},\omega)=
-\frac{1}{\pi}\frac{{\rm Im}\Sigma({\bf k},\omega)}{(\omega+\mu-
\varepsilon_{{\bf k}}-{\rm Re}\Sigma({\bf k},\omega))^2 +
{\rm Im}\Sigma({\bf k},\omega)^2} \label{ak}
\end{eqnarray}
In Fig. \ref{figspect} we show the spectral function obtained from
large-$N$ theory, for an electron at ${\bf k}=(0,0),(1.204,1.204),
(\pi,\pi)$, for different values of $V \rightarrow V_c$. The
spectral density of an electron at the Fermi surface displays a
quasiparticle peak characteristic of a Fermi liquid at
$\omega=\mu$. The rest of spectral weight that is left is
incoherent.

The quasiparticle weight, $Z_{\bf k}$, evaluated at
the Fermi surface is defined as
\begin{equation}
Z_{\bf k}=(1- {\partial {\rm Re}\Sigma({\bf k},\omega)
 \over  \partial \omega} )^{-1}|_{\omega=0}.
\end{equation}

In the inset of Fig. \ref{figspect} we observe how a gradual
suppression of $Z_{\bf k}$ occurs as the charge ordering
transition is approached.  This can be compared to the suppression
of the Drude weight found in Lanczos calculations\cite{Calandra},
which is also evident in the spectral function plotted in Figs.
\ref{Ak0w} and \ref{Akpi2w}.  Spectral weight is transferred from the
quasiparticle peak to the range of energies between $t-3t$, as $V$
tends to $V_c$ due to the scattering of the electrons off the
charge fluctuations associated with short range checkerboard
charge ordering.  The modes close to $(\pi,\pi)$ give the
strongest contribution to the scattering.  The apparent peak
around $\omega=-2t$ should not be interpreted as a quasiparticle
peak but as the lower Hubbard band\cite{Wang1} associated with the
on-site Coulomb repulsion $U$.

\begin{figure}
\includegraphics[height=20cm]{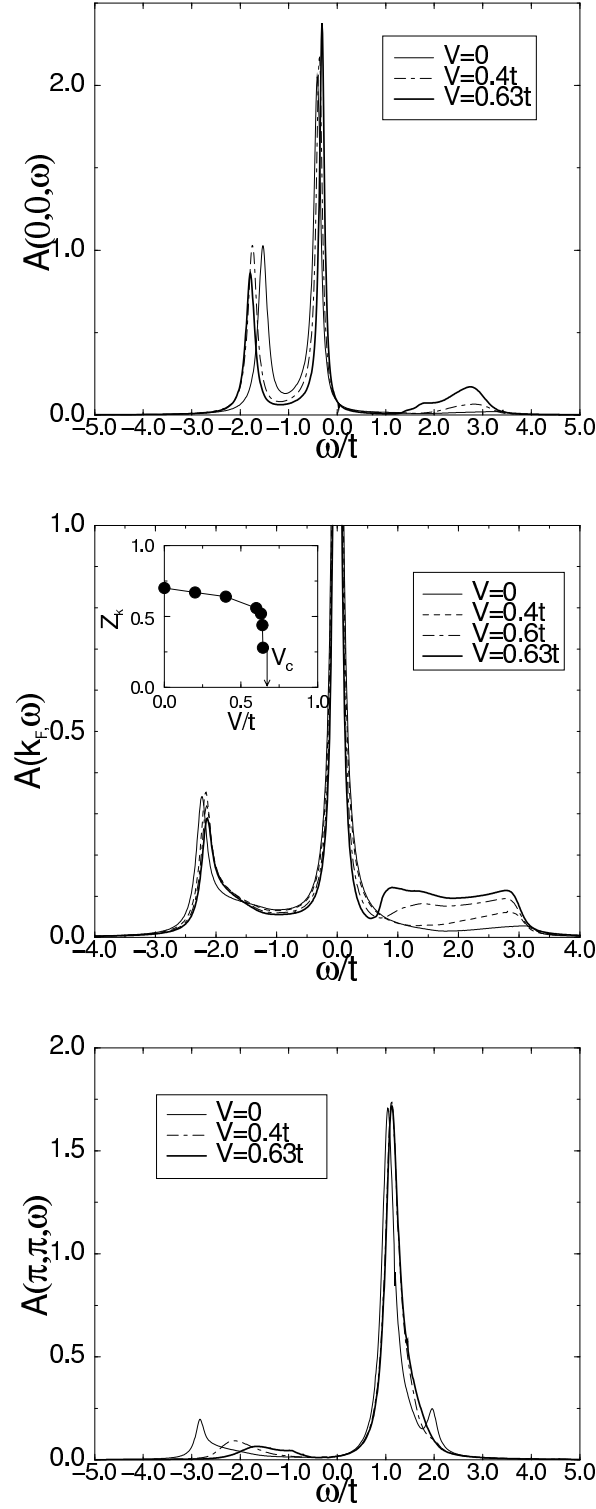}
\caption{Evolution of quasiparticle spectral density of states
computed from large-$N$ theory of an electron at the wavevectors
${\bf k}$=(0,0), $(k_F, k_F)$, and ($\pi,\pi$).  Close to the charge
ordering transition spectral weight is transferred from the
quasiparticle peak to low and intermediate frequencies. The
quasiparticle weight at the Fermi energy, $Z_k$, is rapidly
suppressed (see inset) as the charge ordering transition is
approached $V \rightarrow V_c$. The results presented here can be
compared with $A({\bf k}, \omega)$ computed from Lanczos shown in
Fig. \ref{Ak0w} and \ref{Akpi2w}.} \label{figspect}
\end{figure}

The behavior of the spectral density shown in Fig. \ref{figspect}
can be now understood from the evolution of the real part of the
self energy shown in Fig. \ref{figself}. The scattering of
electrons from the strong charge fluctuations at $(\pi,\pi)$
wavevectors involves large frequencies. This leads to an
enhancement of the real part of the self-energy at large
frequencies which, in turn, produces an increase of spectral
weight at large and intermediate energies. This
behavior is analogous to the one found in metals in the presence
of short range spin fluctuations\cite{Kampf}. Unlike in the case
of nearly antiferromagnetic metals, no new poles induced by the
interaction arise. This is because at quarter-filling no two
points of the Fermi surface are connected by the scattering
wavevector $\bf {q_c}=(\pi,\pi)$, and therefore the effect of the
fluctuations near $\bf{q_c}$ on the electrons is weaker than spin
fluctuations in systems close to half-filling.

In order to test the validity of the large-$N$ approach we have
also computed the spectral densities from Lanczos diagonalization
of finite clusters\cite{Dagotto}
\begin{equation}
A^{(+)}({\bf k},\omega)=\sum_{m}
|\langle m, N_e+1 | c^+_{\bf k \sigma} |0, N_e \rangle|^2
\delta(\omega-(E_m (N_e+1)-E_0(N_e)))
\end{equation}
for adding and electron to the system with $N_e$ electrons and
\begin{equation}
A^{(-)}({\bf k},\omega)=\sum_{m}
|\langle m, N_e-1 | c_{\bf k \sigma} |0, N_e \rangle |^2
\delta(\omega+(E_m (N_e-1)-E_0(N_e)))
\end{equation}
for removing an electron from the $N_e$ electron system.
$E_m$ and $E_0$ denote the excited and ground state
energies of the system and $c^{\dagger}_{{\bf k} \sigma}=1/\sqrt{L}
\sum_j e^{i {\bf k} {\bf R_j} } c^{\dagger}_{j \sigma} $.

\begin{figure}
\includegraphics[width=4.0in,height=4.in]{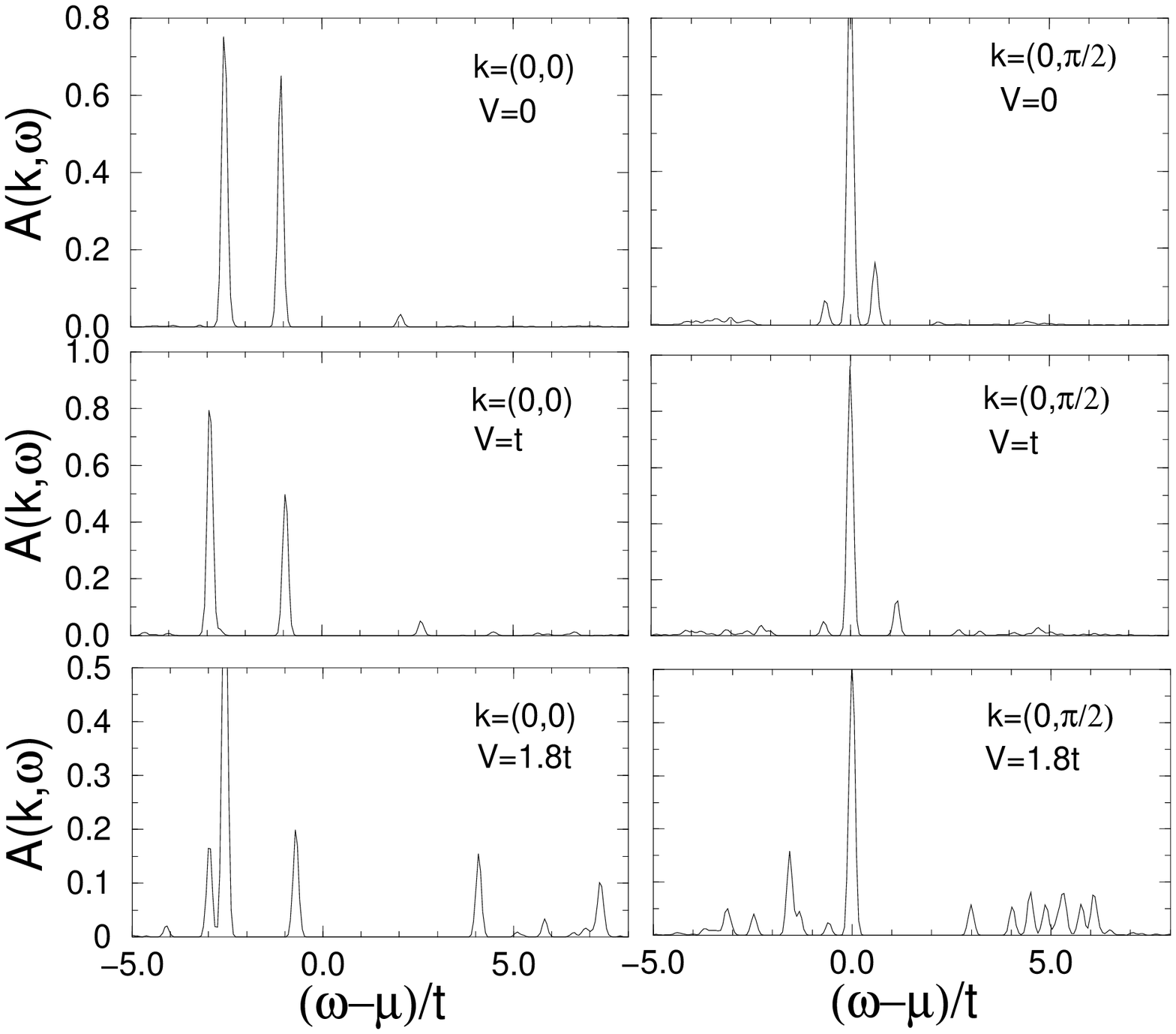}
\caption{Evolution of quasiparticle spectral density of states,
$A({\bf k},\omega)$, at ${\bf k}=(0,0)$ and $(0,\pi/2)$, computed
from Lanczos diagonalisation on a $L=16$ cluster for an extended
Hubbard model at quarter-filling. The on-site Coulomb repulsion is
taken to be $U=20t$ and a broadening of the delta peaks, $\eta
=0.1t$ is used. As the system approaches the metal-insulator
transition, an enhancement of spectral weight at finite
frequencies and a suppression of the weight at the Fermi energy
takes place. At ${\bf k}=(0,0)$, the two sharp peaks are
associated with the lower Hubbard band and the quasiparticle peak.
An overall qualitative agreement with the results from large-$N$
theory is found (see Fig. \ref{figspect}). } \label{Ak0w}
\end{figure}

\begin{figure}
\includegraphics[width=4.0in,height=4.in]{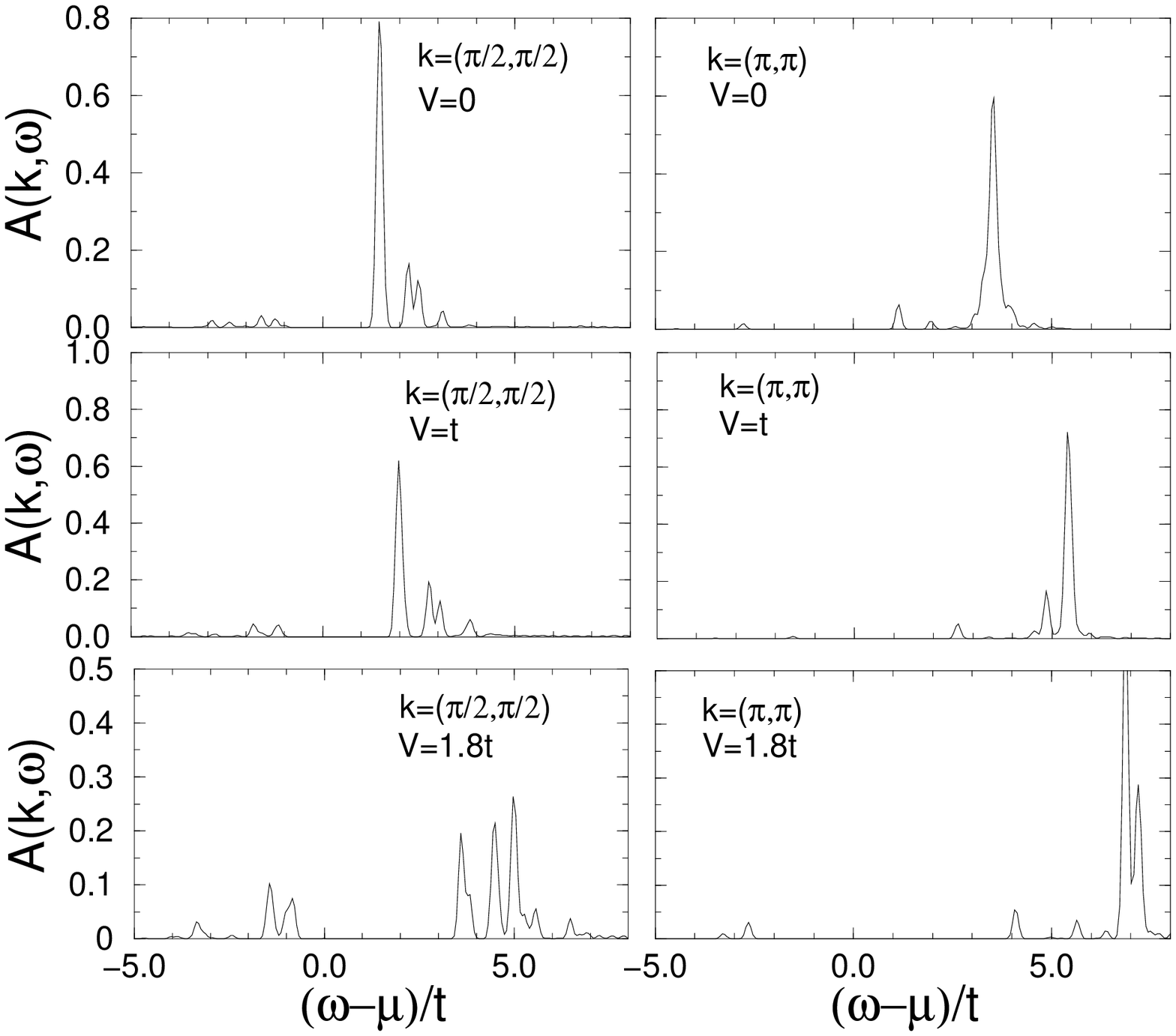}
\caption{Evolution of quasiparticle spectral density of states,
$A({\bf k},\omega)$, at ${\bf k}=(\pi/2,\pi/2)$, and $(\pi,\pi)$,
computed from Lanczos diagonalization on a $L=16$ site cluster for
an extended Hubbard model at quarter-filling.  The on-site Coulomb
repulsion is taken to be $U=20t$ and a Lorentzian broadening of
the delta peaks, $\eta =0.1t$ is used. } \label{Akpi2w}
\end{figure}

In Figs. \ref{Ak0w} and \ref{Akpi2w} we plot the evolution of the
spectral densities calculated with Lanczos techniques for
wavevectors at ${\bf k}=(0,0), (\pi/2,0), (\pi/2,\pi/2)$ and
$(\pi,\pi)$ for different values of $V/t$. At ${\bf k}=(0,0)$ two
sharp peaks are clearly distinguished already for $V=0$. One of
them is the quasiparticle peak and we associate the lower one with
the lower Hubbard band due to the presence of the $U$. For the
nearest wavevectors to the Fermi energy: ${\bf k}=(\pi/2,0)$ and
$(\pi/2,\pi/2)$, we find an enhancement of incoherent spectral
weight at finite frequencies as the charge ordering transition is
approached.

Finally, the total density of states (DOS) can be computed from
\begin{equation}
N(\omega)={1 \over L} \sum_{\bf k} ( A^{(-)}({\bf k},
\omega)+ A^{(+)}({\bf k}, \omega)).
\end{equation}

In Fig. \ref{figdos} we compare the evolution of the DOS,
$N(\omega)$, for increasing $V/t$, calculated with both Lanczos at
$U \rightarrow \infty$ and large-$N$.

\begin{figure}
\includegraphics[width=4.in,height=3.in]{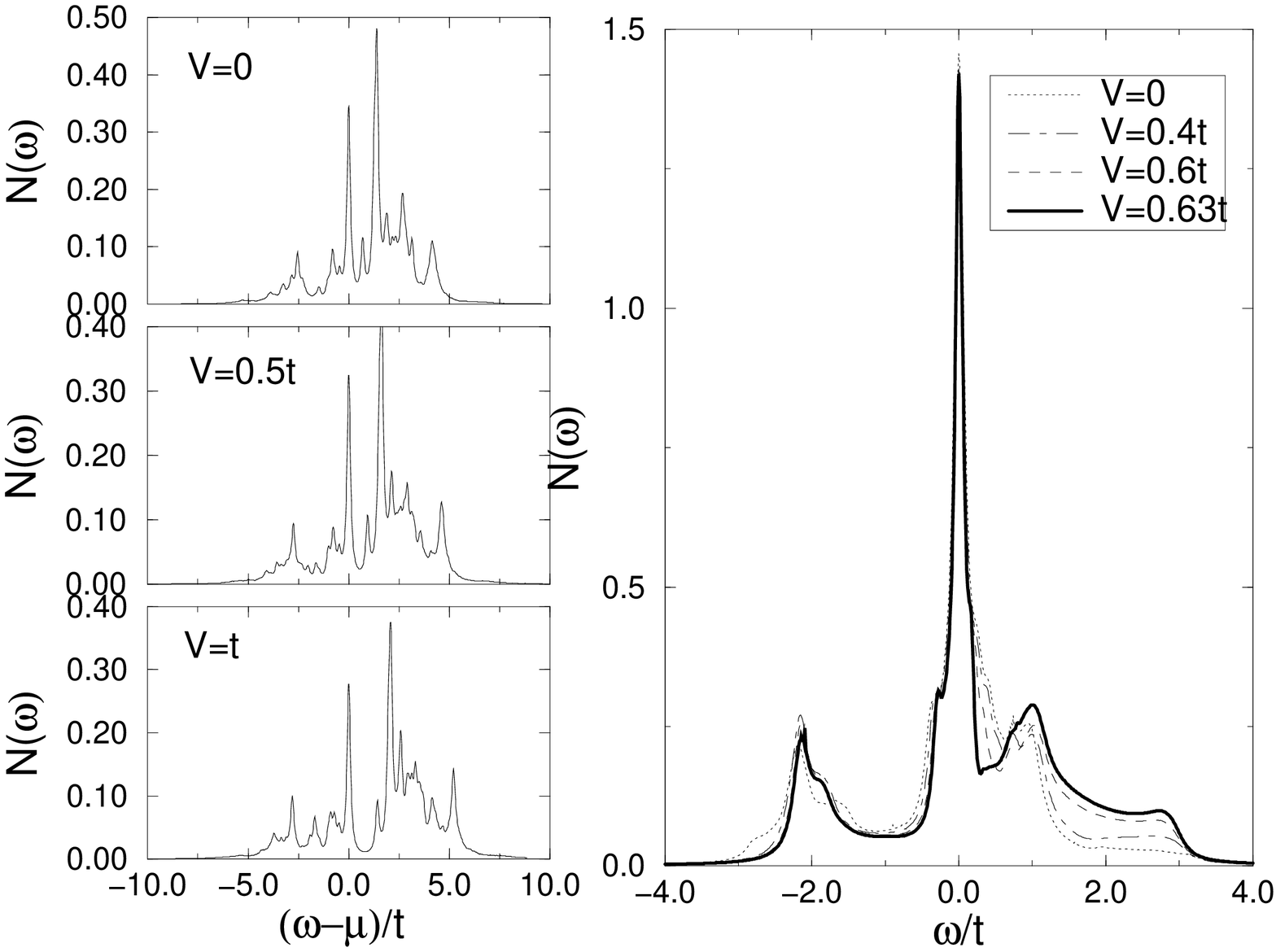}
\caption{Evolution of the total density of states (DOS) as the
charge-ordering transition is approached from the metallic phase.
The left and right panels show results from exact diagonalisation
on a 16 site lattice with $U \rightarrow \infty$ and large-$N$
approaches, respectively. (The critical value of $V$ is $V_c
\approx 2 t$ and $0.65t$, respectively). A Lorentzian broadening
of $\eta=0.1t$ has been introduced in the exact diagonalisation
calculations, to aid comparison with the large N results. As the
intersite Coulomb repulsion $V$ is increased, the density of
states close to the Fermi energy is gradually suppressed
indicating the proximity to the charge ordering transition. At the
same time spectral weight is enhanced for frequencies in the range
$t$ to $3t$ in the large-$N$. The peak at -$2t$ is an incoherent
band associated with the lower Hubbard band. An overall
qualitative agreement between Lanczos and large-$N$ calculations
is found.}
\label{figdos}
\end{figure}
From Lanczos calculations we observe (left panel of Fig.
\ref{figdos}) for $V=0$ a band at about $-3t$, a quasiparticle
band at situated at $\omega=\mu$ and a band running from $t$ to
$5t$. As $V/t$ is increased the weight of the quasiparticle peak
is reduced and weight between $2t$ and $5t$ is gradually enhanced.
Also a suppression of spectral weight at low frequencies occurs as
a precusor effect before the charge ordering transition takes
place. This general behavior is in qualitative agreement with
large-$N$ calculations. Indeed, an incoherent band at negative
frequencies of about $-2t$, associated with the lower Hubbard
band, a suppression of states close to the Fermi energy and an
overall enhancement of spectra between $t$ and $3t$ occurs (see
right panel of Fig. \ref{figdos}). However, we note that the
pseudogap appearing within large-$N$ is less pronounced than in
Lanczos calculations.  This can be attributed to finite size
effects appearing in small cluster Lanczos calculations.

\subsection{Optical conductivity}

It is interesting to analyze the behavior of the optical
conductivity as the system is driven through the charge ordering
transition. Using Lanczos diagonalisation we have computed
\begin{equation}
 \sigma (\omega)=D \delta(\omega) + \frac{\pi e^2}{L} \sum_{n \ne
0}\frac{|\langle n|j_x|0\rangle|^2}{E_n-E_0} \delta( \omega -
E_n +E_0),
\label{opt}
\end{equation}
where $j_x$ is the current in the x-direction, $E_0$ the ground
state energy and $E_n$ the excited energies of the system. $e$ is
the electron charge and the Drude weight is denoted by $D$.

The following sum rule\cite{Maldague} is satisfied by
$\sigma(\omega)$,
\begin{equation}
\int_0^{\infty}  \sigma (\omega) d \omega =  - \frac{\pi e^2}{4 L}
<0| T |0>.
\label{sum}
\end{equation}
where $T$ is the kinetic energy operator,
which is the first term in the Hamiltonian (1).

The optical conductivity is plotted in Fig. \ref{figopt}, for increasing values of the
ratio $V/t$ and fixed $U=20t$. At $V=0$ we
find a Drude peak and a broad mid-infrared band centered at about
$3t$. As $V/t$ is increased the mid-infrared band is enhanced and
a well-defined feature builds up at the lower edge of the mid-infrared
band, at frequencies of about $2t$. Also an incoherent band
present at larger energies of the order of $U$ (not shown for
clarity) is gradually suppressed and its associated weight
transferred to the mid-infrared band as $V$ is increased. From the
behavior of spectral densities and DOS shown in Figs.
\ref{figspect}-
\ref{figdos}, we attribute the
enhancement of optical weight observed in the mid-infrared range to an
increase in the incoherent excitations carried by each
quasiparticle as a result of charge fluctuations associated with
short range checkerboard charge ordering. From the behavior of the
spectral densities, $A({\bf k},\omega)$ shown in Fig. \ref{Ak0w} and \ref{Akpi2w}
and assuming that a lowest order diagram (neglecting
vertex corrections) is enough to compute the optical conductivity we would
attribute the low energy feature to transitions between the incoherent
band carried by each quasiparticle and the quasiparticle peak situated at the Fermi
energy. This interpretation is plausible if one notes that the low energy
feature observed in Fig. \ref{figopt} moves together with the broad band
as $V/t$ is increased. Similar results would be obtained from
large-$N$ theory evaluating the bubble Feynman diagram for the optical
conductivity, as the spectral densities obtained are similar to the
ones obtained from Lanczos diagonalization.

\begin{figure}
\resizebox{!}{4.0in}{\includegraphics{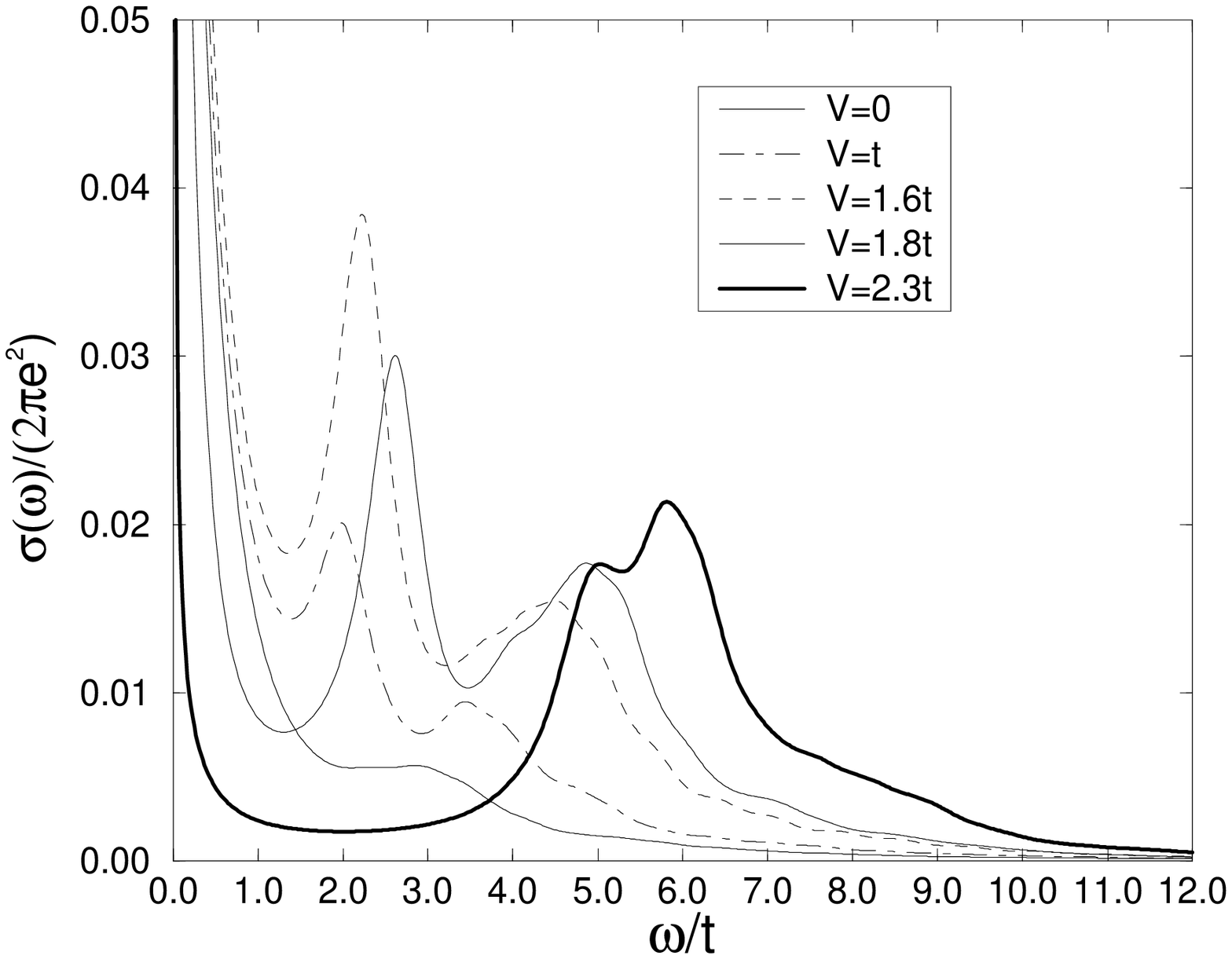}}
\caption{Evolution of the optical conductivity computed from
Lanczos diagonalisation as the system is driven through the
metal-insulator transition. The calculation is performed on a
$L=16$ site cluster, $U=20t$ and different $V$, with a Lorentzian
broadening of $\eta=0.4t$. Enhancement of optical weight at low
frequencies is found as $V$ is increased. The broad band situated
at about $3V$ (for large $V/t$) is due to incoherent transitions
between different sites induced by the intersite Coulomb
repulsion. We interpret the low energy feature appearing at about
$2t$ as a consequence of transitions between the incoherent band
and the quasiparticle peak found in the spectral densities $A({\bf
k}, \omega)$ for wavevectors on the Fermi surface.} \label{figopt}
\end{figure}

\section{Connection to experimental results}
\begin{figure}
\resizebox{!}{4.0in}{\includegraphics{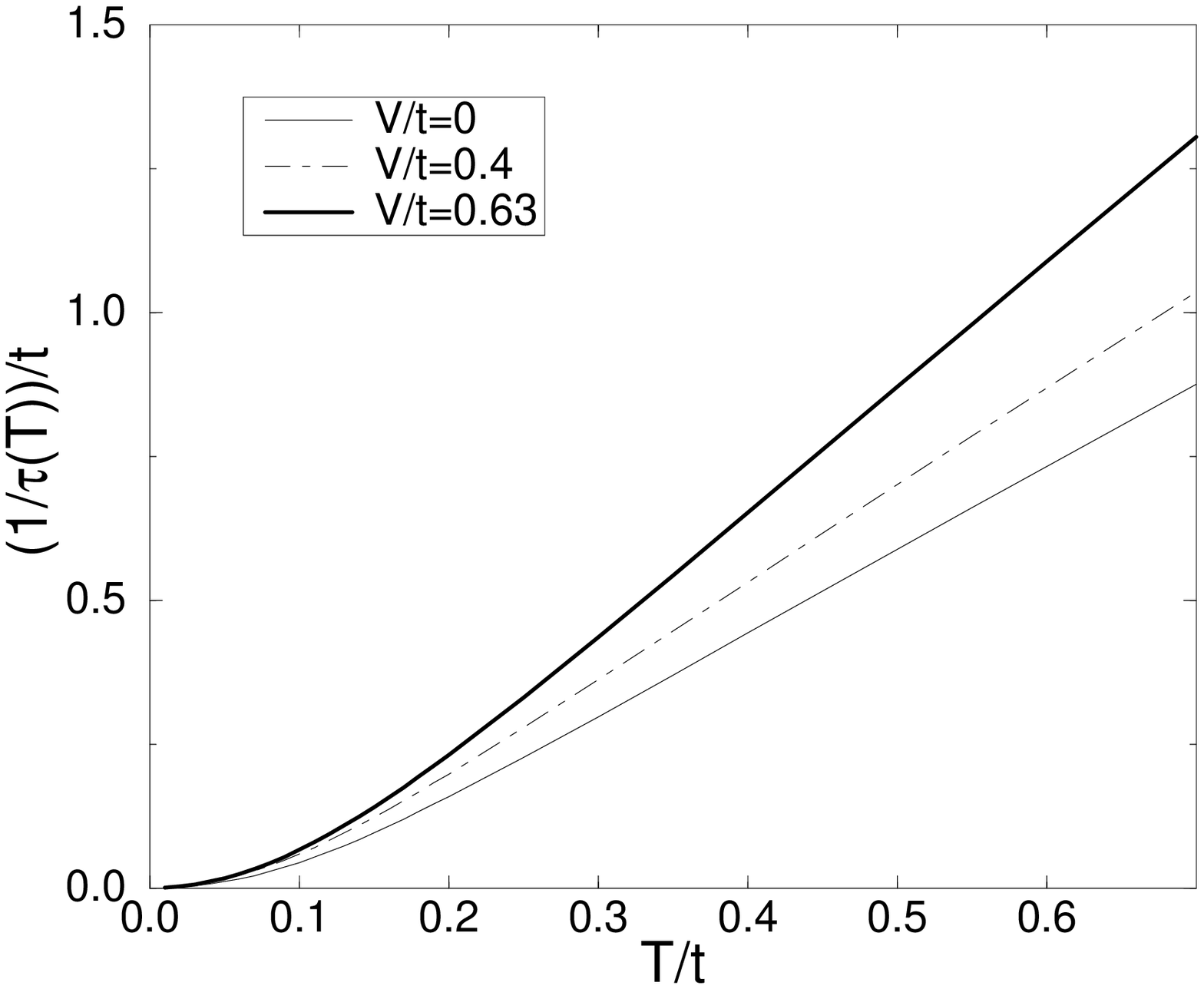}}
\caption{Temperature dependence of the scattering rate
$1/\tau(T)=-2 {\rm Im}\Sigma({\bf k_F},\omega=0)$.
This scattering, which is due to charge fluctuations,
 increases as the charge-ordering transition is approached.
 As the temperature increases above about
$T^* \approx 0.22t$, $1/\tau(T)$ changes from a $T^2$ dependence
to a linear behavior in $T$. This temperature scale depends only
slightly with $V$, so that large-$N$ theory (through O($1/N$))
predicts Fermi liquid behavior close to the charge ordering
transition at temperatures below $T^*$.} \label{figtau}
\end{figure}

Recent experiments with Raman scattering\cite{Yakushi} and optical
conductivity measurements \cite{Ouyang} on the insulating salt
$\theta$-(BDT-TTP)$_2$Cu(NCS)$_2$ find that the checkerboard
charge ordered state discussed in this paper is indeed the ground
state. This gives experimental support to the model discussed
here. A discussion of other possible orderings within more
complicated models can be found in the work by Seo\cite{Seo}
and Clay, Mazumdar, and Campbell\cite{Clay}.

We review now the experiments on resistivity measurements on several
quarter-filled organics,  and make contact with the predictions of
the large-$N$ approach presented.

From the imaginary part of the self-energy (Eq. (\ref{iself})) we
can obtain the behavior of the inverse of the lifetime of the
quasiparticles, $1/\tau(T)=-2{\rm Im }\Sigma(k_F,0)$ with
temperature as shown in Fig. \ref{figtau}. From this plot we
obtain a temperature scale, $T^* \approx 0.22 t$, at which
$1/\tau(T)$, changes from $T^2$ to $T$ behavior.  The temperature
scale defined by $T^*$ decreases only slightly as we approach the
charge ordering transition remaining finite as $V \rightarrow V_c$
(through O($1/N$). This is in contrast to dynamical mean-field
approaches where a similar low temperature scale is suppressed as
the Mott-Hubbard metal insulator (driven by $U$ instead of $V$) is
approached \cite{Georges}. Hence, Fermi liquid behavior is found
below this temperature scale even close to the charge ordering
transition occurring at $V \approx V_c$. Presumably, higher order
corrections in the $1/N$ expansion may suppress the region where
the system behaves as a Fermi liquid as $V \rightarrow V_c$ .
Future work should focus in understanding this issue better.

We have also computed the temperature dependence of the effective
mass defined as: $m^*/m=1/Z_{k}(T)$, evaluated at the Fermi surface
and is shown in Fig. \ref{figmasses}. Large-$N$ theory predicts an increase of $m^*/m$
as the temperature is raised for $V \rightarrow V_c$. This means
that the system becomes more incoherent as the temperature is
increased. Interestingly this behavior is also found in the
Hubbard model in the limit of infinite dimensions close to the
Mott metal-insulator transition\cite{Rozenberg}. However in that
case the system is close to a metal-insulator transition which takes
place between two non-ordered phases, in contrast to the charge ordering transition
discussed here. At $V=0$ the effective mass is temperature
independent as one would expect from a weakly interacting system.
At the lowest temperatures we obtain enhanced effective masses in the range
1.3 to 2, for $V/t$, varying from 0 to 0.63.

\begin{figure}
\resizebox{!}{3.5in}{\includegraphics{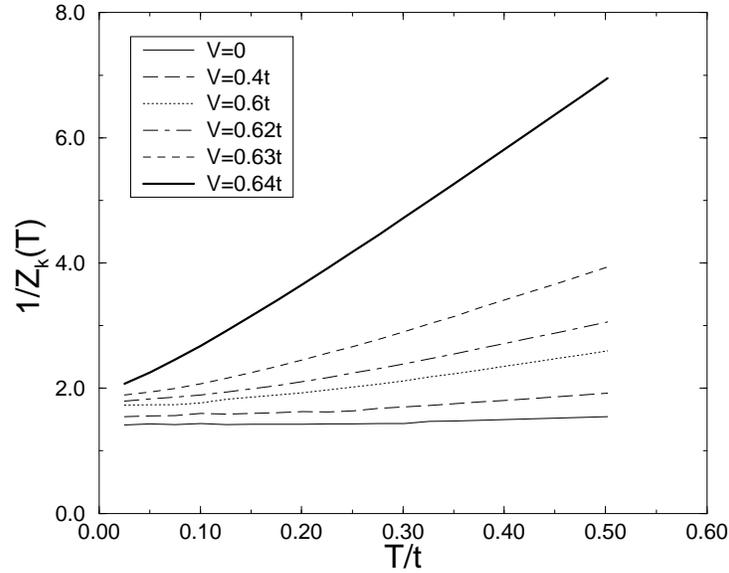}}
\caption{Temperature dependence of the effective mass of an
electron on the Fermi surface, $m^*/m \equiv 1/Z_{k}(T)$, as obtained
from large-$N$ theory through $O(1/N)$. As the system is driven closer to the
charge ordering transition a stronger increase of the effective mass with
$T$ is found.
 } \label{figmasses}
\end{figure}

In Fig. \ref{figoptexp} we show optical conductivity data of
$\theta$-(BEDT-TTF)$_2$CsCo(SCN)$_4$ along the
 a-direction\cite{Yamamoto} at $T=15$ K. This salt is metallic down to
temperatures of about 10 K. At this temperature a charge ordering
transition to an insulating phase takes place.
The observed optical conductivity
displays a  band situated at
1.2 eV, a mid-infrared band appearing at frequencies of about
0.25 eV, and a feature appearing at low frequencies of about
0.13 eV. In the same figure we present a comparison
of our exact diagonalization calculations of the optical
conductivity performed on a 16 site cluster, $U/t=20$ and
$V/t=1.2$ where we have set the hopping energy scale to $t=0.061
eV$, so that we associate the mid-infrared band observed
experimentally with the one from exact diagonalization
calculations. In this way, we recover the main features appearing
in the experimental data, including the incoherent high-frequency feature and
the feature appearing at low frequencies. This behavior is
commonly observed in metallic $\theta$-salts close to the
metal-insulator transition \cite{Calandra} and from Fig.
\ref{figoptexp} we notice that the low energy feature can be
misinterpreted as being part of the Drude peak.  Caution is
in order when comparing our results with experimental
data as shown in Fig. \ref{figoptexp}
because some features like the dip appearing at about 0.17 eV
have been interpreted in $\theta$-(ET)$_2$RbZn(SCN)$_4$ (where
a structural transition takes place with lowering temperature)
as being caused by the coupling to vibronic modes of the ET molecules
\cite{Yamamoto}. More experimental and theoretical work
is needed to understand this issue better.
\begin{figure}
\resizebox{!}{3.5in}{\includegraphics{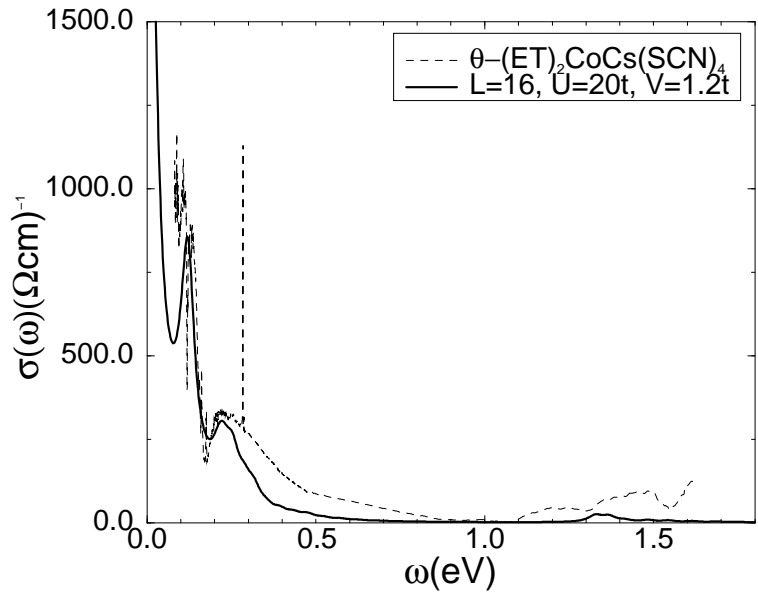}}
\caption{Comparison of the optical conductivity computed from Lanczos 
diagonalisation on $L=16$ site clusters with experimental results \protect{\cite{Yamamoto}} for the
metallic salt $\theta$-(BEDT-TTF)$_2$CsCo(SCN)$_4$. For the exact
diagonalization results we have chosen: $U=20t$ and $V=1.2t$.
In order to fit the data we chose $t=0.061$ eV, which can be compared
to values from H\"uckel band structure calculations \protect{\cite{mori}}.
The lattice parameters for $\theta$-(BEDT-TTF)$_2$CsCo(SCN)$_4$
are  $a=9.804 \AA$, $c=4.873 \AA$
and $V_{cell}=4V_{mol}=2074 \AA^3$, where $V_{cell}$ and $V_{mol}$
are the volumes per unit cell and per molecule, respectively. The
broad band at about 0.25 eV and the low energy feature at 0.13 eV
can be explained from short range charge ordering induced by the
intersite Coulomb repulsion $V$. This behavior is characteristic of
several quarter-filled layered metallic salts which undergo a metal-insulator
transition at sufficiently low temperatures.
}
\label{figoptexp}
\end{figure}

\section{Superconductivity}
\label{sec5}

In the present section we discuss the possibility of
having superconductivity close to the charge ordering transition
induced by the short-range charge fluctuations which appear
in the metallic phase. Here we extend the discussion
presented in Ref.[\onlinecite{prlmerino}] and provide
full details of the calculations.  We also consider the binding energy of holes using
a Lanczos calculation.

\subsection{Large-N: pairing symmetry}

Within the large-$N$ approach, superconductivity is possible at
$O(1/N)$. As we have already seen at $O(1)$, our approach
describes quasiparticle excitations with renormalized masses.
Interaction between these quasiparticles can appear at the
next-to-leading order of $O(1/N)$. The effective interaction
between electrons are those represented diagrammatically in
Fig.\ref{figfeynman}(b); only the three-leg vertex shown in Fig.\ref{figvertex}(a)
contributes to the effective interaction through order $O(1/N)$.

Using the Feynman rules introduced above (see Fig. \ref{figfeynman}(b)), the
interaction between the quasiparticles $\Theta({\bf k, k'})$ reads
\begin{eqnarray}
\Theta({\bf k-k'},\omega_n-\omega_{n'})&=&-[\mu^2
D_{RR}({\bf k-k'},\omega_n-\omega_{n'})+2 \mu
D_{R\lambda}({\bf k-k'},\omega_n-\omega_{n'})
\nonumber \\
&+& D_{\lambda\lambda}({\bf k-k'},\omega_n-\omega_{n'})]
\label{theta}
\end{eqnarray}
where $\mu$ is the chemical potential and $D_{ab}$ are the
components of the boson propagator which are obtained from Dyson's
equation (\ref{Dab}).

In Fig. \ref{figveff} we plot the dependence of $V_{eff}({\bf
q}={\bf k-k'} ) \equiv \Theta({\bf q},\omega \rightarrow 0)/(\delta/2)$,
that is the static limit of the effective interaction mediating
the possible pairing between the quasiparticles. This clearly
shows the development of the singularity due to checkerboard
charge ordering at the $(\pi,\pi)$ wavevectors.
We note that the effective interaction,
 $\Theta$, obtained from Eq.(\ref{theta}),
which is valid on the whole Brillouin zone,
coincides only with the one obtained using slave boson approaches, when
it is evaluated at the Fermi surface \cite{Kotliar}. This is
not true for wavevectors outside the Fermi surface.

\begin{figure}
\resizebox{!}{3.0in}{\includegraphics{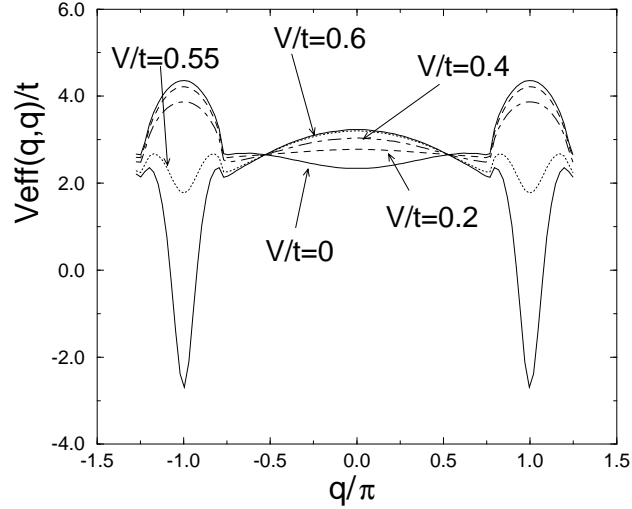}} \caption{Behavior
of the effective potential between quasiparticles,
$V_{eff}(q_x,q_y)$, as a function of momenta for increasing $V/t$
values along the $q_x=q_y=q$ direction. As $V/t \rightarrow
(V/t)_c$, the effective potential becomes negative at the
$(\pi,\pi)$ points, becoming singular at the transition to the
checkerboard charge ordered insulator. It is the momentum
dependence of the potential shown here which leads to the d$_{xy}$
symmetry of the Cooper pairs.
This calculation was done using the large-$N$ approach through $O(1/N)$.} \label{figveff}
\end{figure}

\begin{figure}
\resizebox{!}{3.0in}{\includegraphics{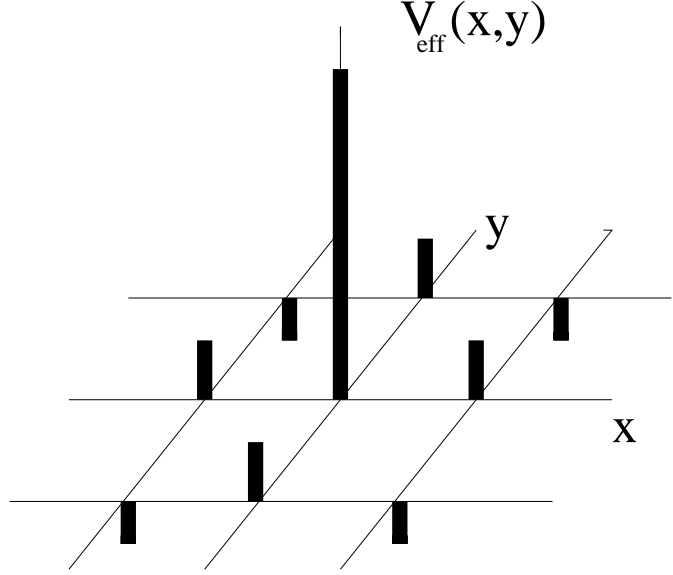}}
\caption{Dependence of the effective couplings with $V/t$ as
defined in Eq. (\ref{lambda}) in the different symmetry channels.
Close to the charge ordering transition pairing in the d$_{xy}$
channel becomes favorable while other possible pairing symmetries
are repulsive for any $V$.} \label{figcoup}
\end{figure}

In weak coupling, we use this effective potential to compute the
effective couplings in the different pairing channels or
irreducible representations of the order parameter, $i$
$(i=(d_{x^2-y^2}, d_{xy}, s))$. In this way we project out the
interaction with a certain symmetry. The critical temperature,
$T_c$, can then be estimated from: $T_{ci}= 1.13 \omega_0
\exp(-{1/|\lambda_i|})$, where $\omega_0$ is a suitable cutoff
frequency and $\lambda_i$ are the effective couplings with
different symmetries. These are defined as
\cite{Scalapino,Kotliar}:

\begin{equation}
\lambda_i={1 \over (2 \pi)^2} {\int (d {\bf k} /|v_{\bf k}|) \int (d {\bf k'}/|v_{\bf k'}|)
g_i({\bf k'})
V_{eff}({\bf k'-k}) g_i({\bf k}) \over \int (d {\bf k}/|v_{\bf k}|)  g_i({\bf k})^2 }
\label{lambda}
\end{equation}
where the functions: $g_i({\bf k})$, encode the different pairing
symmetries, and $v_{\bf k}$ are the quasiparticle velocities at
the Fermi surface. The integrations are restricted to the Fermi
surface. $\lambda_i$ measures the strength of the interaction
between electrons at the Fermi surface in a given symmetry channel
$i$. If $\lambda_i > 0$, electrons are repelled. Hence,
superconductivity is only possible when $\lambda_i <0$.
In Fig. \ref{figcoup} we plot the dependence of the effective
couplings in the possible symmetry channels with $V/t$.
\begin{figure}
\resizebox{!}{3.0in}{\includegraphics{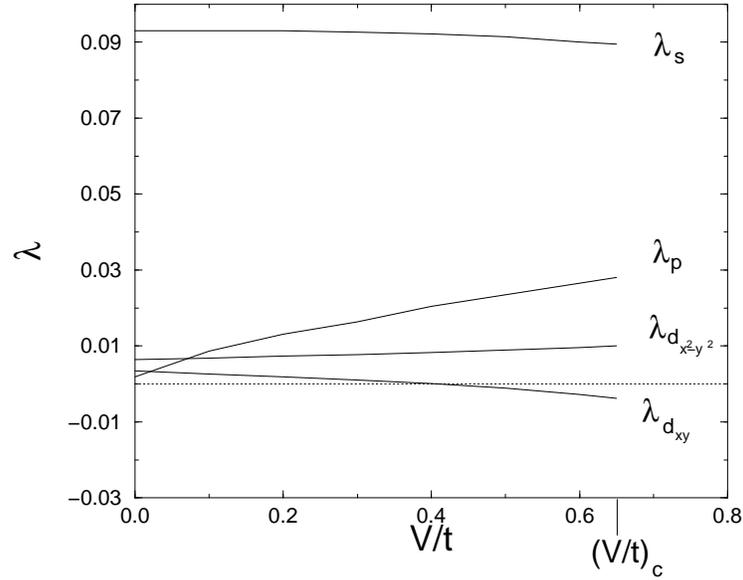}}
\caption{Schematic plot showing the Fourier transform of the
effective potential, $V_{eff}(q_x,q_y)$ for $V/t \approx (V/t)_c $
to real space. $V_{eff}(x,y)$ is understood as follows. A
quasiparticle is placed at the origin. For instance, if another
quasiparticle is placed also at the origin there is a large
repulsion between them due to the large on-site Coulomb repulsion.
This is shown by the large positive vertical bar at the origin. At
neighboring sites (along the x, and y-directions) the effective
potential between quasiparticles is also positive, {\it i.e.},
repulsive. However, at the next-nearest-neighbor sites (along the
diagonals of the lattice), the potential becomes attractive. This
leads to d$_{xy}$ pairing of the quasiparticles. }
\label{figveffxy}
\end{figure}

We observe that near the charge ordering instability but still in
the metallic phase: $V < V_c$, the coupling in the $d_{xy}$
channel, $\lambda_{d_{xy}}$, becomes attractive whereas other
couplings become more repulsive. However, we note that the
couplings are rather small. This implies that critical
temperatures are expected to be small. Similar conclusions have
been reached with large-$N$ treatments of the $U$-infinite Hubbard
model at $V=0$ close to half-filling \cite{Greco}. Exact
diagonalisation results also lead to similar
conclusions\cite{Dagotto}. Although the eigenvalues are small, our
results are nontrivial and show a tendency of the model to Cooper
pairing in the $d_{xy}$ channel. Intuitively one would think that
superconductivity is less and less favorable when increasing $V$,
due to the repulsion between electrons in neighboring sites.
Contrary to this intuition we find that short range charge
fluctuations can mediate pairing close to $V_c$.

In Fig. \ref{figveff} we observe that as we increase $V$, the
effective interaction becomes more repulsive at small momentum
transfer. On the other hand, they become more attractive for
momenta transfer close to $(\pi,\pi)$. That $d_{xy}$ symmetry is
favored can be more clearly understood from Fig. \ref{figveffxy}
which shows a schematic plot of the Fourier transform of
$V_{eff}({\bf q})$ (see Fig. \ref{figveff}). One sees that the
potential is negative for an electron placed at the nearest
neighbor diagonal sites of the lattice while it is positive along
the x and y-directions. This is in contrast to the effective
potential resulting from spin-fluctuations which show the opposite
behavior. This can be understood from previous calculations on a
3-D extended Hubbard model close to half-filling within RPA
performed by Scalapino, Loh and Hirsch \cite{Scalapino}, which
found that the effective potential for charge fluctuations has a
{\it negative} divergence at $(\pi,\pi)$ as the transition is
approached whereas for spin fluctuations it is positively
diverging\cite{Scalapino}.  Due to the fact that the Fermi surface
at one-quarter filling is small (no two points in the Fermi
surface are connected by the $(\pi,\pi)$ wavevector), the
interaction is less effective in inducing pairing as compared to
spin fluctuations in nearly antiferromagnetic metals close to
half-filling.

The $T_c$ values
shown in the phase diagram in Ref. [\onlinecite{prlmerino}] are
larger than the values that would be obtained for $T_{cd_{xy}}$ from
the BCS equation. In Ref. [\onlinecite{prlmerino}]
$T_c$ (for each $V$) was taken to be the temperature below which
the coupling $\lambda_{d_{xy}}$ becomes
negative. Such a calculation
is indicative for superconductivity. However, the appropriate way
to obtain $T_c$ is by solving the associated Eliashberg gap equation.

In conclusion, in the present study we find tendencies to
superconductivity in the $d_{xy}$ channel mediated by short range
charge fluctuations which appear in the metallic
phase close to the charge ordering instability.

\subsection{Lanczos diagonalization: binding energies}

We have computed the binding energy of two holes for different values of
$V/t$ and $U=20t$ on different clusters. The binding energy of two
holes for $L=16$ is defined as\cite{Hirsch,Horsch,Riera}
\begin{equation}
E_B(2 \ holes)=(E(6)-E(8))-2(E(7)-E(8)),
\end{equation}
where $E(N_e)$ is the energy of the system with $N_e$ electrons.

In Fig. \ref{figbind} we plot the binding energy for different
values of $V$.  Initially, as we increase $V$, the binding energy
becomes more positive.  This corresponds to the weak coupling
regime where one naively expects that $V$ keeps the quasiparticles
farther apart.  Further increasing of $V$ closer to the
metal-insulator transition but still in the metallic phase leads
to a negative binding energy of two holes. From finite size
scaling of the binding energy of clusters up to $L=20$ sites, we
find that this happens at about $V \approx 1.6t$.  This finite
size scaling is shown in the inset of Fig. \ref{figbind} for
values of $V$ close to the metal-insulator transition. Remarkably,
the binding energy of two holes changes only slightly in the range
of values $t < V< 2t$ when going from $L=16$ to $L=20$. However,
in the region $V > V_c=2t$, the results change significantly as we
increase the size of the cluster from $L=16$ to $20$, and the
binding energy would eventually extrapolate to a positive value in
the thermodynamic limit.  It is interesting that this region
corresponds to the insulating phase found earlier\cite{Calandra}
from Lanczos calculations of the Drude weight. An interpretation
of our results can be made based on previous works
\cite{Horsch,Hirsch} which studied the binding energy of an
extended Hubbard model of the high-T$_c$ superconductors close to
half-filling as a function of $V$.  As $V$ is increased charge
fluctuations associated with checkerboard charge ordering increase
and the quasiparticles existing at small $V$ gradually dress up
with a cloud of checkerboard charge excitations.  This leaves
signatures in the one-electron dynamical properties as explained
in previous sections.  Further increasing of $V$ leads to pairing
between the quasiparticles mediated by the strong charge
fluctuations \cite{Horsch}. Increasing $V$ even further drives the
system into the insulating phase.

Summarizing, a definitive conclusion about superconductivity
cannot be made from our results. However, it is remarkable that
both large-$N$ and Lanczos diagonalization calculations show a
similar tendency to pairing of quasiparticles in the metallic
phase close to the charge ordering transition. Large-$N$ theory
singles out the d$_{xy}$ symmetry as the preferred pairing channel
of the quasiparticles.  This symmetry is consistent with the
checkerboard charge order present in the region where the Lanczos
pairing energy becomes negative.

\begin{figure}
\resizebox{!}{3.0in}{\includegraphics{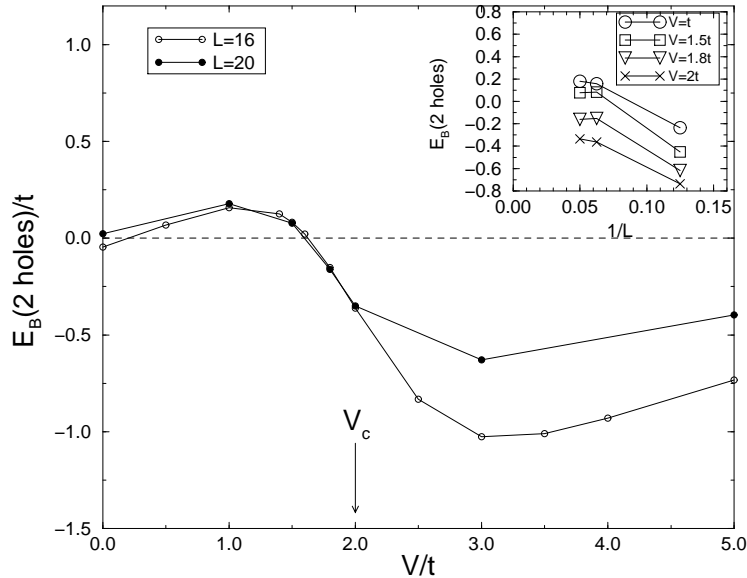}}
\caption{Binding energy of two holes for different values of $V$
and $U=20 t$ from exact diagonalisation calculations on $L=16$ and
$L=20$ clusters. Close to the charge ordering transition, in the
range, $V >1.6t $, binding of two holes becomes favorable. $V_c$
denotes the value of $V$ at which the metal-insulator transition
is estimated to take place from Lanczos calculations of the Drude
weight. Note that the value of $V$ at which E$_B$(2 holes) becomes
negative is robust against increasing the cluster size from $L=16$
to $L=20$. These results are consistent with large-$N$
calculations supporting the possibility of pair formation close to
the charge ordering transition.} \label{figbind}
\end{figure}

\section{Conclusions}

In summary, using a combination of large-$N$ and Lanczos
techniques we have explored the dynamical properties of the
extended Hubbard model at quarter-filling. This is motivated by
its relevance to a large class of superconducting layered organic
molecular crystals. The correlation functions computed from
large-$N$ theory and Lanczos techniques are found to be in good
agreement. Indeed, close to the charge ordering transition driven
by the intersite Coulomb repulsion, $V$, several features are
found: (i) The quasiparticle weight, Z$_{\bf k}$, is rapidly
suppressed near the charge ordering transition. (ii) Spectral
density is enhanced at frequencies ranging from $t$ to $3t$, which
is also reflected in the optical conductivity. (iii) From the
computation of the electron scattering rate we find Fermi liquid
behavior up to $T \approx T^*$, where $T^*$ does not depend
strongly on $V$. For $T > T^*$ the scattering rate behaves linearly
with $T$. (iv) From large-$N$ calculations we find that
superconductivity with d$_{xy}$ symmetry is favored close to the
charge ordering transition. Exact diagonalization calculations of
the binding energy of two holes are consistent with this
possibility.

Given our prediction of unconventional superconductivity
in the $\theta$ and $\beta''$ molecular crystals it
is desirable that more measurements be made to test for this.
The only evidence so far comes from a measurement of
the temperature dependence of the London penetration
depth of $\beta''$-(BEDT-TTF)$_2$SF$_5$CH$_2$CF$_2$SO$_3$.  It
was found to go like $T^3$ at low
temperatures.\cite{prozorov} This is inconsistent
with an s-wave state, but also deviates significantly
from the linear temperature dependence expected
for a d-wave state.
On the other hand, the temperature dependence of the heat capacity
is consistent with s-wave pairing.\cite{Wosnitza2}
Electronic Raman scattering could be used
to investigate the symmetry of the superconducting order
parameter. For $d_{xy}$ symmetry, Raman scattering in the
superconducting state should show, at low frequencies, either
$\omega$, $\omega^3$, or $\omega$ behavior for $B_{1g}$, $B_{2g}$
and, $A_{1g}$ symmetries, respectively \cite{Devereaux}.

An important issue to be resolved concerns the role of spin fluctuations
in the quarter-filled materials.
To first order in 1/$N$, the large-$N$ approach
used here does not take spin fluctuations into account.\cite{Sudbo}
 Measurements of the nuclear magnetic resonance
relaxation rate and Knight shift should be done in the
metallic phase for the relevant superconductors.
If the spin fluctuations are not important
there should be no enhancement of the Korringa ratio.
This is in contrast to the
large enhancements seen in
$\kappa$-(BEDT-TTF)$_2$X superconductors which are
close to an antiferromagnetic Mott insulator \cite{slichter}.

One way to theoretically investigate the role of the antiferromagnetic spin
fluctuations
that may be present near the charge-ordering transition is
as follows. Well into the insulating charge ordered phase
(i.e., for $V \gg t$) it is known that there is an
antiferromagnetic exchange interaction $J' = 4t^4/9V^3$
that acts along the {\it diagonals} of the square lattice\cite{McKenzie}.
Some remnant of this effect will still be present
when there is short-range charge order.
This could be modelled by considering a $t-J'-V$ model
where the $J'$ acts only along the diagonals.
This model could be studied by the same large-$N$ method
used previously to study a large family of $t-J-V$ models\cite{Vojta}.
There it was found that the superexchange, acting along
the vertical and horizontal lattice directions, produced $d_{x^2-y^2}$
superconductivity.  Based on that work we anticipate that the effect of the
superexchange, which now acts in
directions rotated by 45 degrees,
will be to produce $d_{xy}$ superconductivity.
Hence, it is possible that charge and spin fluctuations
work together co-operatively to produce $d_{xy}$ pairing.

\acknowledgments

We acknowledge helpful discussions with J. S. Brooks, E. Dagotto,
M. Dressel, A. Foussats, R.  Giannetta, P. Horsch, E. Koch, R.
Noack, B. Powell, J. Wosnitza, Z.  Hasan, J. Riera,  R. Zeyher and M. Vojta.
We thank J. Wosnitza for showing us unpublished experimental
results and K. Yamamoto for sending his unpublished optical data
to us. J. M. and M.C. were supported by two Marie Curie
Fellowships of the European Community program ``Improving Human
Potential'' under Contract No. HPMF-CT-2000-00870 (J.M.) and No.
IHP-HPMF-CT-2001-01185 (M.C.). Work  at UQ was supported  by the
Australian Research Council. A. G. thanks to Fundaci\'on Antorchas
for partial financial support.


\begin{thebibliography}{99}
\bibitem{mori} This is about half the value of 106 meV found in
Reference \protect\onlinecite{Mori}.
However, one should be cautious about taking the H\"uckel and
extended H\"uckel values as definitive. Based on
comparison with results from the local density
approximation it is argued in Reference
\protect\onlinecite{merino}
that the H\"uckel methods tend to systematically
underestimate the hopping integrals.
\end{thebibliography}
\end{document}